\documentclass[fleqn,usenatbib]{mnras}

\usepackage{newtxtext,newtxmath}
\usepackage{subcaption}
\usepackage[T1]{fontenc}
\usepackage{hyperref}

\DeclareRobustCommand{\VAN}[3]{#2}
\let\VANthebibliography\thebibliography
\def\thebibliography{\DeclareRobustCommand{\VAN}[3]{##3}\VANthebibliography}

\usepackage{subcaption}
\usepackage{enumitem}
\usepackage{amsmath}
\usepackage{dsfont}
\usepackage{makecell}
\usepackage{tabularx}
\usepackage{pbox}
\usepackage{bbding}
\usepackage{chngpage}
\usepackage[flushleft]{threeparttable}
\usepackage{booktabs,caption}
\usepackage{mathtools}
\usepackage{floatrow}
\usepackage{multirow}
\usepackage{comment}
\usepackage{xspace}

\newcommand{\hcMpc}{\mathrm{h^{-1} cMpc}}
\newcommand{\sml}{R_{\mathrm{sm}}}
\newcommand{\lth}{\lambda_{\mathrm{th}}}
\newcommand{\fgas}{\ensuremath{f_{\mathrm{gas}}}}
\newcommand{\lr}{(\lambda_{\mathrm{th}}, R_{\mathrm{sm}})}

\newcommand{\IGM}{\ensuremath{\mathrm{IGM}}}
\newcommand{\figm}{\ensuremath{f_\IGM}}

\newcommand{\Afrac}{\frac{X_\mathrm{H} m_\mathrm{He}+2Y_\mathrm{He} m_\mathrm{H}}{m_\mathrm{H} m_\mathrm{He}} \Omega_b \rho_c(z)}

\newcommand{\simba}{\mbox{{\sc Simba}}\xspace}

\newcommand{\ff}{\ensuremath{\simba\mathrm{-50}}}
\newcommand{\nx}{\ensuremath{\simba\mathrm{-nox}}}
\newcommand{\nj}{\ensuremath{\simba\mathrm{-nojet}}}
\newcommand{\nagn}{\ensuremath{\simba\mathrm{-noagn}}}
\newcommand{\nf}{\ensuremath{\simba\mathrm{-nofb}}}

\newcommand{\intdelm}{\ensuremath{\int(1+\delta_m)ds}}

\usepackage{xcolor}
\usepackage[dvipsnames]{xcolor} %to use color names
\title[Effect of Feedback on Cosmic Web]{\simba Simulation: The Effect of Feedback Physics\\ on Matter Distribution in the Cosmic Web}

\author[Chenze Dong et al.]{
Chenze Dong,$^{1,2}$\thanks{E-mail: dong-chenze@g.ecc.u-tokyo.ac.jp}
Florian Dedieu,$^{3}$
Daniela Galárraga-Espinosa,$^{2}$
Khee-Gan Lee,$^{1,2}$ 
Daniele Sorini,$^{4}$
\newauthor{
Romeel Davé$^{5,6}$}
\\
$^{1}$Center for Data-Driven Discovery, Kavli IPMU (WPI), UTIAS, The University of Tokyo, Kashiwa, Chiba 277-8583, Japan
\\
$^{2}$Kavli Institute for the Physics and Mathematics of the Universe, The University of Tokyo, 5-1-5 Kashiwanoha, Kashiwa, Chiba, 277-8583, Japan
\\
$^{3}$ENS Paris-Saclay, Paris-Saclay University, 4 Av. des Sciences, 91190 Gif-sur-Yvette, France
\\
$^{4}$Institute for Computational Cosmology, Department of Physics, Durham University, South Road, Durham, DH1 3LE, United Kingdom
\\
$^{5}$Institute for Astronomy, University of Edinburgh, Royal Observatory, Edinburgh EH9 3HJ, UK
\\
$^{6}$Department of Physics and Astronomy, University of the Western Cape, Bellville, Cape Town 7535, South Africa
}

\date{Accepted XXX. Received YYY; in original form ZZZ}

\pubyear{\the\year{}}

\begin{document}
\label{firstpage}
\pagerange{\pageref{firstpage}--\pageref{lastpage}}
\maketitle

% Abstract of the paper
\begin{abstract}
The discrepancy between the early-time estimation and late-time observation of the cosmic baryon content -- the `missing baryon problem' -- is a longstanding problem in cosmology.
Although recent studies with fast radio bursts (FRBs) have largely addressed this discrepancy, the precise spatial distribution of these baryons remains uncertain due to the effects of galaxy feedback. 
Cosmological hydrodynamical simulations such as \simba have shown that the partitioning of baryons between the intergalactic medium (IGM) and haloes is sensitive to feedback models, motivating the connection between baryon distribution and feedback physics.
With the \simba simulation suite, this study investigates how feedback affects the distribution of matter within large-scale cosmic structures, with implications for FRB foreground modelling.
We apply the T-web method to classify the cosmic web into different structures: knots, filaments, sheets, and voids. 
We then analyse how the different feedback variants of \simba affect the distribution of matter within each structure.
Our results show that in \simba, the fractions of IGM gas in different cosmic web structures vary by only a few percent under different feedback models. 
However, jet feedback produces noticeable changes in the gas distribution within structures, enhancing the diffuse IGM on the outskirts of filaments and knots, potentially biassing the empirical relation between matter distribution and dispersion measure.
This research provides a new perspective on the impact of feedback on the IGM and motivates a refined data model for FRB foreground mapping.
\end{abstract}

% Select between one and six entries from the list of approved keywords.
% Don't make up new ones.

%%%%%%%%%%%%%%%%%%%%%%%%%%%%%%%%%%%%%%%%%%%%%%%%%%

%%%%%%%%%%%%%%%%% BODY OF PAPER %%%%%%%%%%%%%%%%%%

\section{Introduction} \label{sec:intro}
%\subsection{Missing Baryon Problem}
The ``missing baryon problem'' describes the observational deficit of the cosmic baryon density parameter, $\Omega_b$, observed in the low-redshift universe when compared to early-time measurements from cosmic microwave background (CMB) studies (e.g., \citealt{Planck_2016, Planck_Collab:2020}) as well as Big Bang Nucleosynthesis constraints from light element abundances \citep{Steigman:2010,Cooke:2024}.
\cite{Fukugita:1998} noted that a mere 10\% of the universe's baryonic matter is found in galaxies as stars or interstellar medium (ISM) gas, with the majority existing in a state of diffuse, ionised intergalactic medium well outside of observed galaxies. 

This issue was extensively studied in both theory and observation in the 2000s, as exemplified in the works by  \cite{Fukugita:2004, Cen:2006, Bregman:2007}.
The pioneering simulation study conducted by \cite{Cen:1999} identified the warm-hot intergalactic medium (WHIM) as a potential reservoir for the missing baryons. 
Detecting the WHIM presents a significant challenge since it is not sufficiently hot to emit X-rays significantly and not sufficiently cool to allow for radio emissions.
Subsequent works (e.g., \citealt{Kang:2005, Dave:2001, Kravtsov:2002, Kang:2007, Tornatore:2010, Hong:2014}) revealed that the production of such WHIM was related to so-called ``baryonic effects'', especially gas shock heating from gravitational collapse and galactic feedback.

Simultaneously, various observational efforts are made to locate the missing baryons. 
Guided by predictions from simulations \citep{Cen:2001, Yoshikawa:2003, Cen:2006a, Hallman:2007, Dave:2010, Oppenheimer:2012}, Lyman-$\alpha$ absorption and OVI absorption have been utilised in attempts to detect the presence of the WHIM \citep{Fang:2002, Bregman:2007, Prochaska:2017, Stocke:2013, Nicastro:2018, Kovacs:2019}, alongside the use of the Sunyaev-Zel'dovich (SZ) effect \citep{Shao:2011, Ferraro:2016, Chiu:2018, deGraaff:2019}. 
Despite these comprehensive observational efforts, by 2019, 20\% of baryons remained undetected.

The advent of fast radio bursts (FRBs; see reviews \citealt{Cordes:2019, Glowacki:2024}) as extragalactic probes has significantly altered the situation. 
FRBs are radio pulses lasting milliseconds, originating from extragalactic sources with unknown mechanisms (see \citealt{Platts:2019} for a living catalogue of FRB theories). 
The dispersion measure (DM) of an FRB,
\begin{equation}\label{eq:define_dm}
\mathrm{DM} = \int n_{e} \frac{\mathrm{d}s}{1+z},    
\end{equation}
quantifies its frequency-dependent time delay, determined by the integration of the electron density $n_e$ along the line-of-sight path element $\mathrm{d}s$.
Given that $\approx 99\%$ of extragalactic baryons are ionised after cosmic reionization \citep[e.g.,][]{Neeleman:2016}, the DM is effectively a volume-weighted probe of cosmic baryons, in contrast to most other probes of cosmic baryons, which are halo-centric. 
By analysing 7 FRBs with host galaxy identification (``localised'') detected by ASKAP, \cite{Macquart:2020} reported that the redshift-DM relationship of these FRBs was in good agreement with standard $\Lambda$CDM cosmology.
This result has technically `detected’ the missing baryons but still does not address the question of where they reside, as the FRB DM measures the integrated electron density along the line-of-sight.
Then the relative distribution of cosmic baryons in different parts of the Universe, connected to the scatter of the redshift-DM relation of FRBs, remains unclear.

In modern cosmological hydrodynamical simulations (e.g., \citealt{Schaye:2010,Vogelsberger:2014, Hopkins:2014, Bryan:2014, Schaye:2015, McCarthy:2017, Pillepich:2018, Dav__2019}), galaxy `feedback processes' such as energy injection from stellar winds, supernova explosions, and AGN-driven outflows are introduced primarily to regulate excessive star formation. 
Feedback has been found to have a substantial impact on the overall distribution of baryonic gas, especially in the WHIM \citep{Haider:2016, Martizzi_2019}. 
Furthermore, different feedback models predict different partitions of baryons within, around, and beyond galactic haloes (\citealt{Heckman:2017,Davies:2019, Tollet:2019,Christiansen:2020,  Davies:2020,Lim:2021, Sorini2022, Ayromlou:2023, Angelinelli:2023, Appleby:2023, Sorini:2024, Sorini:2025}) and across distinct cosmic web environments \citep[e.g.][]{Martizzi_2019, Bradley:2022}. 
Thus, measuring the mass fraction of baryons across different scales and environments represents a potentially interesting method for constraining the physics of feedback processes.

Although the baryon distribution imprinted in the FRB DM is a promising probe of feedback, the relatively low number of localized FRBs limited the application of this approach. 
Even with many efforts to detect FRBs by Canadian Hydrogen Intensity Mapping Experiment (CHIME; \citealt{CHIME:2021}), Commensal Real-Time ASKAP Fast Transient (CRAFT; \citealt{Macquart:2010}), MeerKAT TRAnsients and
Pulsars (MeerTRAP; \citealt{Sanidas:2018}), and Deep Synoptic Array (DSA; \citealt{Kocz:2019}), the localized FRBs to date are far from enough ($\sim 10^3$ are needed according to \citealt{Batten:2022}) to constrain feedback models.
However, \cite{Lee_2022} argued that combining FRBs dispersion measures with spectroscopic observations of foreground galaxies (``FRB foreground mapping'') enhances the precision of joint constraints on the intergalactic medium and circumgalactic medium (CGM) baryon distributions beyond that of localized FRBs alone. 
An early attempt by \cite{Simha:2020} involved decomposing DM observed from a single FRB into contributions from the Milky Way, the extragalactic cosmic baryons, and the host galaxy. 
This idea was further extended in the FRB foreground mapping survey, FLIMFLAM, (\citealt{Lee_2022, Khrykin:2024flimflam, Huang:2024}), where the extragalactic baryon contribution is further divided into the contribution from the intervening haloes and the contribution from the IGM. 
Following the integral form of DM in Equation~\ref{eq:define_dm}, one can decompose DM into different origins' contribution.
\begin{equation} \label{eq:DM_FLIMFLAM}
    \mathrm{DM} = \mathrm{DM}_{\mathrm{MW}}+\mathrm{DM}_{\mathrm{haloes}}+\mathrm{DM}_{\IGM}+\mathrm{DM}_{\mathrm{host}}
\end{equation}
where $\mathrm{DM}_{\mathrm{MW}}$ and $\mathrm{DM}_{\mathrm{host}}$ are the contributions of the Milky Way and the FRB host, respectively.
The remaining part, known as ``cosmic baryon contribution'', is further separated into $\mathrm{DM}_{\mathrm{haloes}}$ (circumgalactic medium, CGM, in the intervening haloes) and $\mathrm{DM}_{\IGM}$ (intergalactic medium).
The decomposition improved the efficiency of investigating the cosmic baryon distribution by incorporating additional observations of foreground galaxies, even with a limited FRB sample (See Appendix \ref{appendix:FLIMFLAM_intro} for more detail).

In a comparative study of several feedback variants of the \simba cosmological hydrodynamical simulation, \citet{Khrykin:2024} found that the overall balance of cosmic baryons between the halo CGM and large-scale diffuse IGM is significantly altered by the feedback model adopted.
In particular, they found that the feedback from long-range AGN jets can increase the IGM gas fraction by approximately 20\% and decrease the CGM baryon fraction by around 50\% relative to weaker feedback models.
Their results motivate the parametrisation of $\mathrm{DM_{cosmic}} = \mathrm{DM_{IGM}} + \mathrm{DM_{haloes}}$ in FLIMFLAM as a probe of feedback, although \citet{Hussaini:2025} recently claimed evidence of feedback from lower limits on $f_\mathrm{igm}$ in low-DM localised FRBs from the DSA survey.

% In this study, we follow up on \citet{Khrykin:2024} 
With this in mind, the first step we need to address in this paper is to explore whether different feedback models might redistribute cosmic baryons across the different parts of the cosmic web. 
Specifically, we employ the T-web method to classify $z=0$ cosmic web (e.g., \citealt{Hahn_2007, Forero_Romero_2009, Cautun:2014}) in the \simba simulations into ‘voids’, ‘sheets’, ‘filaments’, and ‘knots’, based on the local dynamics of density field collapse. 

Subsequently, we measure the impact of the feedback on baryon partition from a cosmic web perspective and validate the feasibility of incorporating the cosmic web information in FRB foreground mapping analysis through
\begin{equation}\label{eq:new_decomp}
    \mathrm{DM}_{\mathrm{IGM}} = \mathrm{DM}_{\mathrm{voids}} + \mathrm{DM}_{\mathrm{sheets}} + \mathrm{DM}_{\mathrm{filaments}} + \mathrm{DM}_{\mathrm{knots}}
\end{equation}
In parallel, we examine the `local' baryon-to-matter ratio $\fgas$ within the framework of cosmic web classification and discuss how this can be used to refine FRB DM modelling.
Throughout this paper, we restrict our analysis to the baryon distribution at $z=0$; we will leave the exploration of the direct connection with the FRB observables and the situation at higher redshifts to future works.

The paper is structured as follows: in Section~\ref{sec:methods}, we present the \simba simulation utilised in this study and introduce the T-web method for cosmic web classification. 
In Section~\ref{sec:cls}, we then demonstrate our classification applied to the \simba{} suites.
We compare the partition of matter in different IGM environments in Section~\ref{sec:matter_partition} to verify the improvement on the refined model (Equation~\ref{eq:new_decomp}; details in Appendix \ref{appendix:FLIMFLAM_intro}).
Then, we examine the correlation between overdensity and gas density in Section~\ref{sec:f_gas}. 
In Section~\ref{sec:discussion}, we discuss the results and implications for future observations. 
Finally, we summarise our main findings in Section~\ref{sec:conclusion}. 
We assume the Planck15 cosmology \citep{Planck_2016} ($\Omega_m = 0.3$, $\Omega_{\Lambda} = 1 - \Omega_m = 0.7$, $\Omega_b = 0.048$, $h = 0.68$, $\sigma_8 = 0.82$, $n_s = 0.97$) to stay consistent with the \simba simulation. 

\section{Data and Methods} \label{sec:methods}

In this section, we give a brief introduction to the simulation data used in Section~\ref{subsec:Simba}, and the method of defining the cosmic structures in Section~\ref{subsec:class}. 
We also show how we map the particle-based simulation data into the density fields in Section~\ref{subsec:distribution}. 

\subsection{The \simba simulation} \label{subsec:Simba}
In this study, we adopt the \simba simulation \footnote{http://{Simba}.roe.ac.uk/} \citep{Dav__2019} for our cosmic web analysis.
\simba is a modern suite of galaxy formation simulations, a successor to the \texttt{Mufasa} simulation \citep{Dav__2016}, which aims to reproduce the physics of the universe at diverse spatial scales in a full cosmological context. 
In the following section, we introduce the specifications of \simba and its feedback variants.

\subsubsection{Cosmology and Numerical Scheme}

\simba adopts a $\Lambda$CDM cosmology following the Planck15 measurements \citep{Planck_2016}.
The simulation is performed with the massively-parallel gravito-hydrodynamic code \texttt{GIZMO}, in its Meshless Finite Mass version.
Gas elements and dark matter (DM) are simulated together, the latter being represented through a set of collisionless Lagrangian particles solved via a tree-particle-mesh algorithm, and shocks are computed via a Riemann solver without the need for artificial viscosity. 

\subsubsection{Gas Physics and Star Formation}
In \simba, the Grackle-3.1 library \citep{Smith_2016} handles all the cooling and heating processes (mainly radiation and photo-ionisation), which take into account metal cooling and the non-equilibrium evolution of the primordial elements. 
The ionisation background is taken from \citet{Haardt_2012}, modified in order to take into account the self-shielding effect in a self-consistent way, following the prescription of \citet{Rahmati_2013}.

Star formation is modelled with a \citet{Schmidt_1959} law for $\mathrm{H_2}$, %and where $\mathrm{H_2}$
which is assessed on the basis of the subgrid model of \citet{Krumholz_2011}, modified to take into account variations in numerical evolution \citep{Dav__2016}. 
Star formation rate (SFR) is evaluated from the $\mathrm{H_2}$ density and the dynamical time only \citep{Dav__2019, Kennicutt_Jr__1998}. Metal enrichment takes into consideration 11 elements ($\mathrm{H}$, $\mathrm{He}$, $\mathrm{C}$, $\mathrm{N}$, $\mathrm{O}$, $\mathrm{Ne}$, $\mathrm{Mg}$, $\mathrm{Si}$, $\mathrm{S}$, $\mathrm{Ca}$, $\mathrm{Fe}$) which are introduced by enrichment processes including Type $\mathrm{Ia}$ supernovae ($\mathrm{Ia}$SNe) \citep{Iwamoto_1999}, Type $\mathrm{II}$ supernovae ($\mathrm{II}$SNe) \citep{Nomoto_2006}, and Asymptotic Giant Branch (AGB) stars \citep{Oppenheimer_2006}, as well as the fact that a part of the metal can be contained in dust. 

\subsubsection{Stellar and SMBH Feedback}\label{subsec:feedback_prescription}
The model of star formation-driven winds used by \simba involves decoupled metal-enriched kinetic winds, whereby 30$\%$ of the wind particles are ejected in a hot phase, with temperatures determined by the difference between the supernova energy and the winds kinetic energy, while the others are ejected with a temperature of around $10^3\mathrm{\ K}$. 
The mass loading factor and the wind velocity are the main free parameters, chosen in this simulation to be in agreement with the FIRE zoom-in simulations \citep{Muratov_2015, Angl_s_Alc_zar_2017}. See \citet{Dav__2019} for the complete details. The outflows are enriched in metals, accounting for the impact of the $\mathrm{II}$SNe that extricate heavy elements from the ISM. Once ejected, wind particles cannot radiate and are compelled to thermally transfer their energy to the CGM.

\simba also implements supermassive black hole (SMBH) growth and SMBH feedback. 
SMBH growth is modelled differently depending on the physical state of the accreted gas. 
The ISM gas within the SMBH kernel, always cold in this simulation ($T \leq 10^5$ K), follows a torque-limited accretion mode taken from \citet{Hopkins_2011}, which is physically related to the gravitational instabilities of cold gaseous discs that drive mass inflows. 
The non-ISM hot gas ($T \geq 10^5$ K) usually has a more spherical distribution, which explains the choice of a Bondi accretion rate, given by the standard \citet{Bondi_1952} formula shown in Equation~\ref{eq:Bondi_acc}.

\begin{equation}
    \Dot{M}_{\mathrm{Bondi}} = \epsilon_{\mathrm{m}}\frac{4\pi G^2 M^2_\mathrm{SMBH}\rho}{(v^2 + c_s^2)^{3/2}}
    \label{eq:Bondi_acc}
\end{equation}

where $\rho$ is the mean density of the hot gas calculated within the SMBH kernel, $c_s$ is its average sound velocity, and $v$ is its average velocity relative to the centre of the SMBH. 
$\epsilon_{\mathrm{m}}$ is an attenuation factor set such that $\epsilon_{\mathrm{m}}=0.1$ makes the Bondi accretion rate consistent with the previous torque-limited accretion mode \citep{Dav__2019}.

In \simba, there are three implementations of AGN feedback, designed to be consistent with the observed dichotomy in real AGN feedback (see, e.g., \citealt{Heckman_2014}), which usually involves a radiative mode at high accretion rates and a jet mode for slow-accreting AGN. 
First, the fast-accreting AGN can emit what are called \textit{Radiative AGN winds}. This ``high-accretion feedback mode'' is adopted when the SMBH has an Eddington ratio $f_\mathrm{ED} = \dot{M}/\dot{M}_{\mathrm{ED}}\geq 0.2$ and consists of ejecting gas particles from the vicinity of the SMBH along a direction parallel to the total angular momentum of the accretion disc (typically the 256 closest gas particles to the BH). 
The resulting bipolar outflow velocity depends only on the SMBH mass and follows Equation~\ref{eq:RW-velo}, based on the analysis of the X-ray emission of ionised gas stemming from $z < 0.8$ SDSS AGN \citep{Perna_2017a}.

\begin{align}
    v_{\mathrm{RW}}= 500 + \frac{500}{3}\left( \log \left(\frac{M_\mathrm{BH}}{\mathrm{M_\odot}}\right)-6\right)\mathrm{km/s}
    \label{eq:RW-velo}
\end{align}

The ejection of the particles is adiabatic, and their temperature is consistent with that of the electrons observed in actual ionised gas outflows, with an order of magnitude of $10^4\mathrm{\ K}$ \citep{Perna_2017b}.

The ``slow-accreting mode'' is activated when $f_\mathrm{ED} \leq 0.2$. It involves the same bipolar mass-loaded ionised gas ejections as for the high accretion mode, but with an outflow velocity that additionally depends on $f_\mathrm{ED}$. 
This new dependence, denoted by the equation~\ref{eq:jets-velo}, allows for dramatic outflow velocities when the Eddington ratio decreases, with an order of magnitude as high as $10^4\ \mathrm{km/s}$. However, the velocity boost is capped at $7000\ \mathrm{km/s}$, which generally corresponds to $f_\mathrm{ED} \lesssim 0.02$.

\begin{align}
    v_{\mathrm{jets}} = v_{\mathrm{RW}} + 7000 \log \left( \frac{0.2}{f_\mathrm{ED}}\right) \mathrm{km/s}
    \label{eq:jets-velo}
\end{align}

In addition to a low Eddington ratio, the jet mode requires the condition $M_{\mathrm{BH}} >M_{\mathrm{BH,lim}}$, with $M_{\mathrm{BH,lim}} = 10^{7.5}\ \mathrm{M_{\odot}}$, which aims to prevent AGN with low masses from emitting jets when they temporarily accrete little gas, as observed \citep{Bari_i__2017}.

In either case, radiative winds and jets outflows are ejected bipolarly, with an opening angle assumed to be null and a finite cross section extent generally $\lesssim 1\mathrm{\ kpc}$. 
On the one hand, this choice makes sense for the low accretion mode, as AGN jets are actually observed as thin, collimated ejections. 
As for radiative AGN winds, this model may be a bit reductive, as the opening angle is likely to be greater. 
Nevertheless, \citet{Dav__2019} justifies the choice of bipolar ejection by arguing that re-collimation occurs on scales unresolved in \simba.

Finally, fast-accreting SMBH with fully boosted jets can exert \textit{X-ray heating} feedback if the galaxy is poor in gas, with a gas mass fraction $f_{\mathrm{m,gas}} = M_{\mathrm{tot}}/M_{*} < 0.2$. This definition is motivated by the observed inclination of the slow radiative winds to appear in more gaseous galaxies, where there would be important radiative losses \citep{Best_2012}. 
The X-ray heating is applied to the gas particles within the SMBH accretion kernel, computed following Equation~12 of \citet{Choi_2012}, scaling on the inverse of the square of their distance from the SMBH. 
The energy conversion is made depending on whether the gas belongs to the ISM or not. 
The cold ISM gas is simply heated and its temperature is increased depending on the heating flux at its position. Half of the heating energy given to the non-ISM gas is converted into kinetic energy, in the form of a radial outward kick, while the rest is kept as thermal energy. 
The X-ray heating feedback has little effect on the galaxy mass function but plays a critical role for galaxy quenching.

\subsubsection{\simba feedback variants}

The \simba simulation suites contain one flagship run with a volume of $(100 \mathrm{\ cMpc/h})^3$ and full feedback physics, along with many other variant runs that include modifications to volume and physics. To study the properties of the cosmic web affected by different feedback mechanisms, we utilised several $(\mathrm{50\ cMpc/h})^3$ feedback variants of \simba in this research (see Table~\ref{table:data_spec}).
It should be noted that all the \simba feedback variants begin with identical initial conditions (IC), facilitating a field-level comparison among the runs.
We also adopted the \texttt{Caesar}\footnote{https://caesar.readthedocs.io/en/latest/} halo catalogues provided in the \simba data repository.
Most post-processing work on simulation snapshots is done with the lightweight analysis package for \texttt{Gadget} output, \texttt{pygad} \citep{pygad}.

\begin{table*}
\caption{Specification of \simba runs used in this study. All the feedback variants have the same box size and mass resolution, but with different feedback mechanism turning off.
$N_p$ is total number of dark matter particles and gas cells. The $m_{DM}$ and $m_{gas}$ are the mass resolution for dark matter particles and gas cells, respectively.
Each feedback variant run of \simba contains $512^3$ dark matter particles and the same number of gas particles. The detail of feedback implementation is introduced in Section \ref{subsec:Simba}.}
\label{table:data_spec} 
\centering     
\scalebox{0.9}{
\begin{tabular}{ccccccccc}
Simulation & Box size (cMpc/h) & $N_p$ & $m_{\mathrm{DM}} (10^7 M_\odot)$ & $m_{\mathrm{gas}}(10^7 M_\odot) $ & Stellar & AGN winds & AGN jets & AGN X-ray \\ \hline
\multicolumn{1}{l}{Feedback variants} & \multicolumn{1}{l}{} &  &  &  &  &  &  &  \\ \hline
\ff & 50 & $2 \times 512^3$ & 9.6 & 1.82 & Y & Y & Y & Y \\
\nx & 50 & $2 \times 512^3$ & 9.6 & 1.82 & Y & Y & Y & N \\
\nj & 50 & $2 \times 512^3$ & 9.6 & 1.82 & Y & Y & N & N \\
\nagn & 50 & $2 \times 512^3$ & 9.6 & 1.82 & Y & N & N & N \\
\nf & 50 & $2 \times 512^3$ & 9.6 & 1.82 & N & N & N & N
\end{tabular}
}
\end{table*}

\subsection{Classifying the cosmic web} \label{subsec:class}

In this work, we derive the cosmic web classification based on the T-web methodology. 
The latter relies on the three eigenvalues $\lambda_1 \ge \lambda_2 \ge \lambda_3$ of the deformation tensor $\psi_{ij}$, where $\psi_{ij}$ is the Hessian matrix of the rescaled gravitational potential $\tilde\phi = \phi / 4\pi G \bar{\rho}_m$, which also follows Poisson's Equation:
\begin{equation}\label{eq:dt_x}
    \left\{
    \begin{aligned}
    \Delta \tilde{\phi} (\mathbf{x}) &= \delta_m(\mathbf{x})\\
    \psi_{ij}(\mathbf{x}) &= \frac{\partial^2 \tilde{\phi}}{\partial x_i \partial x_j}(\mathbf{x})
    \end{aligned}
    \right.
    \end{equation}
where $\delta_m(\mathbf{x}) = (\rho_m-\bar{\rho}_m)/\bar{\rho}_m(\mathbf{x})$ is the total overdensity. 
To mitigate numerical artefacts on small scales, we applied a Gaussian smoothing kernel to the density field before the subsequent calculation. 
The smoothing scale $\sml$ is also a hyperparameter of the T-web method and defines the scale of features to be extracted in the cosmic web analysis. We choose $\sml = 2\rm\ Mpc/h$, which we will justify later on. 

After the density smoothing step, we calculate the deformation tensor through Fast Fourier Transform (FFT) and then obtain the classification based on the number of eigenvalues larger than a certain threshold $\lth$, which is another hyperparameter in the method.
\begin{equation}\label{eq:classification}
    \mathrm{Classification} = 
    \left\{
             \begin{aligned}  
             &\mathrm{Node}, &\lambda_1 \ge \lambda_2 \ge \lambda_3 \ge \lth\\  
             &\mathrm{Filament}, &\lambda_1 \ge \lambda_2 \ge \lth \ge \lambda_3\\  
             &\mathrm{Sheet}, &\lambda_1 \ge \lth \ge \lambda_2 \ge \lambda_3\\  
             &\mathrm{Void}, &\lth \ge \lambda_1 \ge \lambda_2 \ge \lambda_3\\  
             \end{aligned}
    \right.  
\end{equation}

According to \citet{Hahn_2007}, $\lth$ was set to zero based on the Zel'dovich approximation. 
However, \cite{Forero_Romero_2009} indicated that such a choice is inadequate for filtering structures with large collapsing timescales (i.e., positive but small $\lth$). 

In this study, we selected $(\lth, \sml) = (0.16, 2\,\hcMpc)$ after careful assessment of different combinations of hyperparameters (see Appendix \ref{appendix:convergence_test} for more details). 
We start by choosing a smoothing length of $2$ cMpc/h, which is motivated by the spatial resolution of the density field reconstruction in FLIMFLAM (1.875 cMpc/h). 
This smoothing scale is at the limit of the best resolution that can be achieved when reconstructing density fields using galaxy surveys, which is due to limitations in the observations and density field modelling \citep{Khrykin:2024flimflam}.

For the choice of the eigenvalue threshold, we use $\lth=0.16$, since this choice effectively restores the visual representation of smoothed cosmic web structures. 
At the same time, it aligns the mass and volume fractions of our cosmic web classification with the \cite{Martizzi_2019} to $2\%$ level.

A concise description of our classification workflow is provided below.
\begin{enumerate}[leftmargin=*]
    \item Starting with simulation snapshots, we first run a cloud-in-cell (\texttt{CiC}) algorithm to obtain the matter density field $\rho_m(\mathbf{x})$. 
    The algorithm will loop over all types of simulation particles and distribute the mass of the particles onto a grid representing the simulation volume. 
    We further calculate the overdensity field $\delta_m(\mathbf{x})$ with $0.1\,\hcMpc$ resolution and subsequently smooth $\delta_m(\mathbf{x})$ with a Gaussian kernel with $\sml = 2\,\hcMpc$. \\
    \item The deformation tensor is given by equation~\ref{eq:dt_x};
    here we switched to the Fourier $\mathbf{k}$-space to perform the computation. 
    One can calculate the deformation tensor in Fourier space by
    \begin{equation}\label{eq:dt_k}
        \psi_{ij}(\mathbf{k}) = -k_i k_j \phi(\mathbf{k}) = \frac{k_ik_j}{||\mathbf{k}||^2} \delta_m(\mathbf{k})
    \end{equation}
    Therefore, we obtain the $\mathbf{k}$-space overdensity $\delta_m(\mathbf{k})$ via the Fast Fourier Transform (FFT) and then derive the $\mathbf{k}$-space deformation tensor $\psi_{ij}(\mathbf{k})$ using Equation~\ref{eq:dt_k}. 
    Finally, we perform inverse FFT on $\psi_{ij}(\mathbf{k})$ to get $\psi_{ij}(\mathbf{x})$.\\
    \item For the deformation tensor associated with each voxel, we calculate its eigenvalues $\lambda_1, \lambda_2, \lambda_3$ and obtain the T-web classification from the criteria~\ref{eq:classification} by adopting $\lth = 0.16$. \\
\end{enumerate}

\subsection{Describing the matter distribution} \label{subsec:distribution}

Motivated by FRBs as a probe of baryon distribution, this study aims primarily to explore the impact of feedback models on the distribution of gas from a cosmic web perspective.
The principal methodology involves the volumetric classification detailed in Section~\ref{subsec:class}.
% We consider the distributions of different species: matter (dark matter and baryons), IGM gas, and free electrons. 
We analyse the spatial distributions of several related quantities: total matter (dark matter and baryons), baryonic gas, and free electrons.
% While the matter distribution defines the cosmic web, the latter two are more observationally motivated. 
Of these quantities, the total matter defines the cosmic web, baryonic gas serves as the direct recipient of feedback effects, and electron density represents the most direct observable within FRB measurements.

% We generate the grid of matter density, $\rho_m$, and baryon density, $\rho_b$, by running the \texttt{CiC} algorithm over the mass of simulation particles (dark matter, gas, stars, and black holes) in the snapshots of \simba. 
The electron number $N_{e, \mathrm{cell}}$ of each gas cell is calculated using the mass of the gas cell $M_{\mathrm{gas}}$, the hydrogen abundance $X_H$, and the electron abundance $\eta_e$:  
\begin{equation}
    N_{e, \mathrm{cell}} = \eta_e \frac{X_\mathrm{H}\mathrm{M_{gas}}}{m_\mathrm{p}}
\end{equation}

We generate the grid of matter density $\rho_m$, baryon density $\rho_b$, and electron number $N_e$ by running the \texttt{CiC} algorithm over simulation particles (dark matter, stars, and black holes) and cells (gas) in the $z=0$ snapshot of \simba. 
% Next, we also run the \texttt{CiC} algorithm on $N_e$ to derive the grid of electron distribution. 
We denote the densities of matter, gas, and free electrons as $\rho_m, \rho_g$ and $\rho_e$. The normalised over-densities are $\delta_m = \rho_m / \bar{\rho}_m - 1, \delta_g = \rho_g / \bar{\rho}_g - 1$ and $\delta_e = \rho_e / \bar{\rho}_e - 1$, respectively. 

Combining the density grids and the cosmic web classification, it is straightforward to calculate the mass fractions:

\begin{equation}\label{eq:frac_mass}
    F^{i}_{j} =     \frac{M^{i}_{j}}{\sum_j M^{i}_{j}}
\end{equation}

where the index $i~\in$~\{\ff, \nx, \nj, \nagn, \nf\}, $j \in \{\mathrm{voids, sheets, filaments, knots}\}$. 

Unlike dark matter, the quantities of baryons, gas, and free electrons are subject to change across different feedback scenarios, as feedback mechanisms influence star formation and black hole accretion, leading to variations in baryonic matter remaining as a diffuse medium. 
To remove variations in total amounts, we compute the fraction using the subsequent normalisation:
\begin{equation}\label{eq:norm}
\begin{aligned}
    F^{i}_{j, \mathrm{baryon}} &= \frac{M^{i}_{j, \mathrm{baryon}}}{\sum_j M^{i}_{j, \mathrm{baryon}}},\\
    F^{i}_{j, \IGM} &= \frac{M^{i}_{j, IGM}}{\sum_j M^{i}_{j, \IGM}},\\
    F^{i}_{j, e} &= \frac{N^{i}_{j, e}}{\sum_j N^{i}_{j, e}}.
\end{aligned}
\end{equation}

The gridded density field also allows us to define a `local' gas fraction $\fgas$
\begin{equation}\label{eq:f_gas}
    \fgas = %\frac{\Omega_m}{\Omega_b}
    \frac{\rho_{\mathrm{gas}}}{f_b \rho_{\mathrm{tot}}},
\end{equation}
where $f_b = \Omega_b / \Omega_m$ is the cosmic baryon fraction and $\rho_{\mathrm{gas}}$ is the density value of a given grid voxel.
Thereby, $\fgas>1$ corresponds to a gas excess with respect to the average cosmic gas fraction, while $\fgas<1$ corresponds to a gas deficit.
The definition is inspired by \cite{Khrykin:2024}, where a similar quantity is used to describe the gas content inside haloes. To suppress numerical artefacts while avoiding the loss of the structure of interest when we compute the local gas fraction, we apply minimal Gaussian smoothing $r_{\mathrm{smoothing}}=0.1$ cMpc/h, which corresponds to the resolution of the density grid. 
We select this minimal smoothing scale, rather than the 2 cMpc/h previously used for the cosmic web classification, because $f_{gas}$ only appears in the FRB DM computation, whose resolution is not limited by foreground galaxy observations and density reconstruction. 

\begin{figure*}
    \centering
    \scalebox{0.32}{\includegraphics{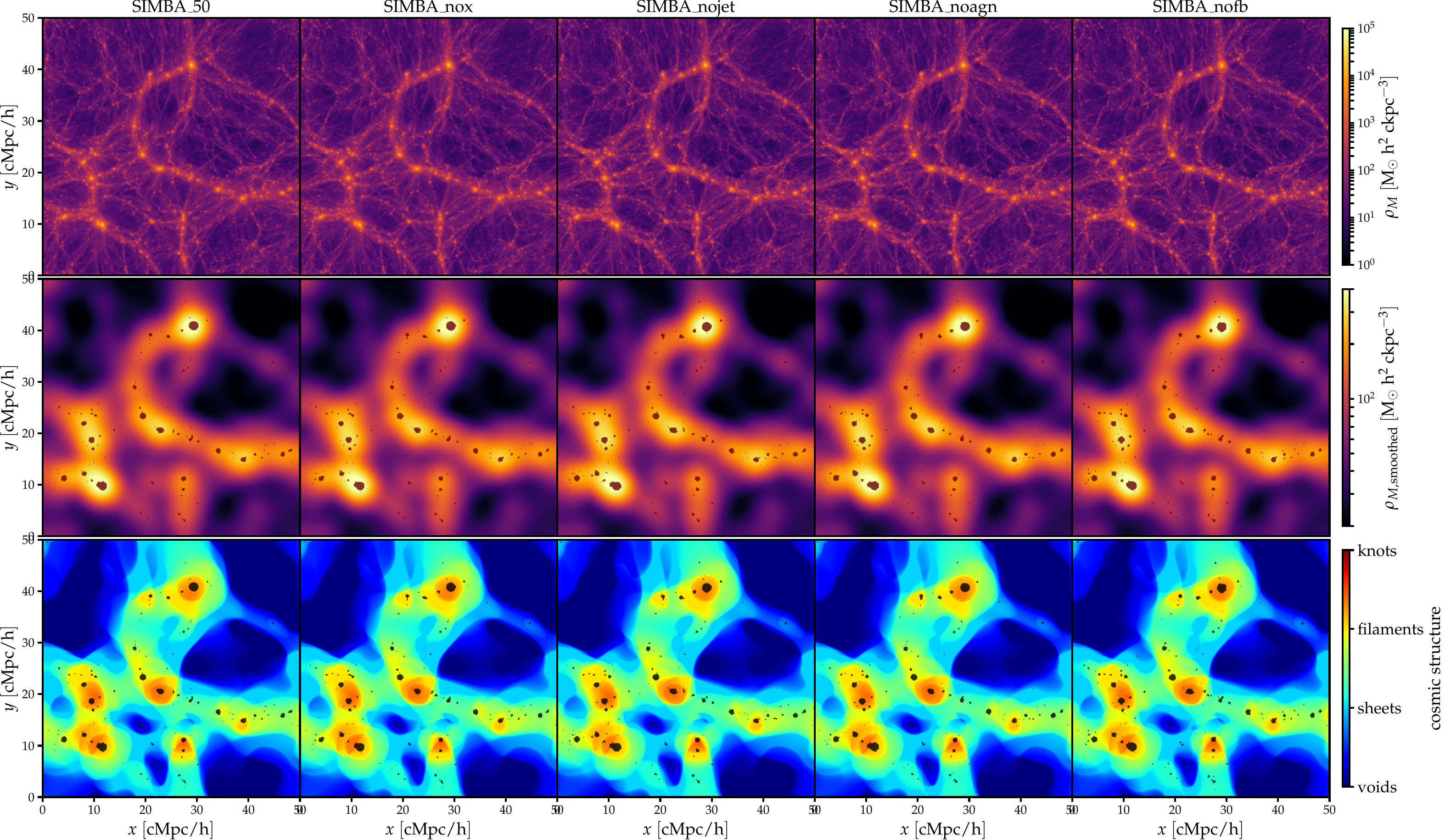}}
    \caption{\textbf{Left to right}: Slices of \simba feedback variants (\ff, \nx, \nj, \nagn, \nf) with 10 cMpc/h thickness showing the full density field (dark matter and baryons) and associated cosmic web classifications. 
    \textbf{Top row}: The slice of density field taken from the $z=0$ snapshot. 
    The ``cosmic web'' is defined with the interconnected filament and the nodes, surrounded by sheets and voids. 
    \textbf{Middle row}: The same slice but smoothed with a Gaussian kernel with a size of $2$ cMpc/h. 
    \textbf{Bottom row}: corresponding averaged classification field, performed with $\lth = 0.16$. Haloes are overplotted as masks in the middle (dark red) and right column (black). Visually, there is little change on the shape of structures among these variants. 
    }
    \label{fig:class_fields}
\end{figure*}

\subsection{Processing the Halo Catalogue} 
We also examined whether the matter, baryons, or electrons are linked to a halo. 
We adopted the \texttt{Caesar} halo catalogue provided by the official \simba repository. 
To eliminate nonphysical haloes caused by numerical artefacts, we filtered the minimum stellar mass with a threshold of $3.64\times 10^8 h^{-1}M_{\odot}$, equivalent to 20 times the average mass of a gas cell. 
For all the feedback variants, the filtered catalogues have a maximum halo mass of about $3\times10^{14} h^{-1}M_{\odot}$.  \ff{}, \nx{} and \nj{} also have similar median halo masses ($2\times10^{11} h^{-1}M_{\odot}$) and minimum halo masses ($6\times 10^{9} h^{-1}M_{\odot}$).
\nagn{} run has a higher minimum halo mass of $1.1 \times 10^{10} h^{-1}M_{\odot}$, and \nf{} run has a lower median halo mass of $4\times10^{10} h^{-1}M_{\odot}$. 
Utilising the $r_{200}$ measurements from the \texttt{Caesar} catalogue, volumetric halo masks for each feedback variant were constructed in parallel with the cosmic web classification mask. 
We define the IGM as the medium beyond the $r_{200}$ masks.
These halo masks were subsequently employed for the calculation of matter distribution within haloes and for visualisation.

\section{Cosmic web classification in feedback variants} \label{sec:cls}

The top panels of Figure~\ref{fig:class_fields} show an example $z=0$ slice of the matter density field in the \simba feedback variants. 
Across the different feedback prescriptions, we visually find no significant differences in the full density field (top row). 
Given that the initial conditions are identical, the similarity of large-scale structures among the feedback variants is not surprising. 
At the same time, the similarity hints at the insensitivity of the large-scale density field to the feedback physics. 

The matter density field smoothed with a 2 cMpc/h Gaussian kernel and the resulting cosmic web classifications are shown in the middle and bottom panels of Figure~\ref{fig:class_fields}. 
The latter are also indistinguishable among the feedback variants, suggesting that the effect of feedback on the cosmic web is subtle when computed directly from the full matter density field (see also, e.g., \citealt{Sunseri_2023}). 
We attribute the subtlety to the fact that the cosmic web 
structure at these large scales is primarily driven by gravitational interactions, which are dominated by dark matter that is largely insensitive to baryonic feedback effects. 
Our smoothing with $2$ cMpc/h Gaussian kernel also contributes to decreasing the apparent impact on baryonic physics.
In Figure~\ref{fig:class_fields}, we also overplot the position of dark-matter haloes on both panels. 
Even though the density and associated classification are similar, we note that the positions of dark-matter haloes  differ slightly from one prescription to another, which may arise from the sub-Mpc impact of the feedback physics or numerical artefacts in the halo finder.

\section{Matter Partition in cosmic environments}\label{sec:matter_partition}
This section illustrates the distribution of matter between the IGM and haloes under the cosmic web classification. 
For specific fraction numbers, refer to Appendix~\ref{appendix:data_table}.

\begin{figure*}
    \centering
    \begin{subfigure}[b]{0.45\textwidth}
        \scalebox{0.35}{\includegraphics{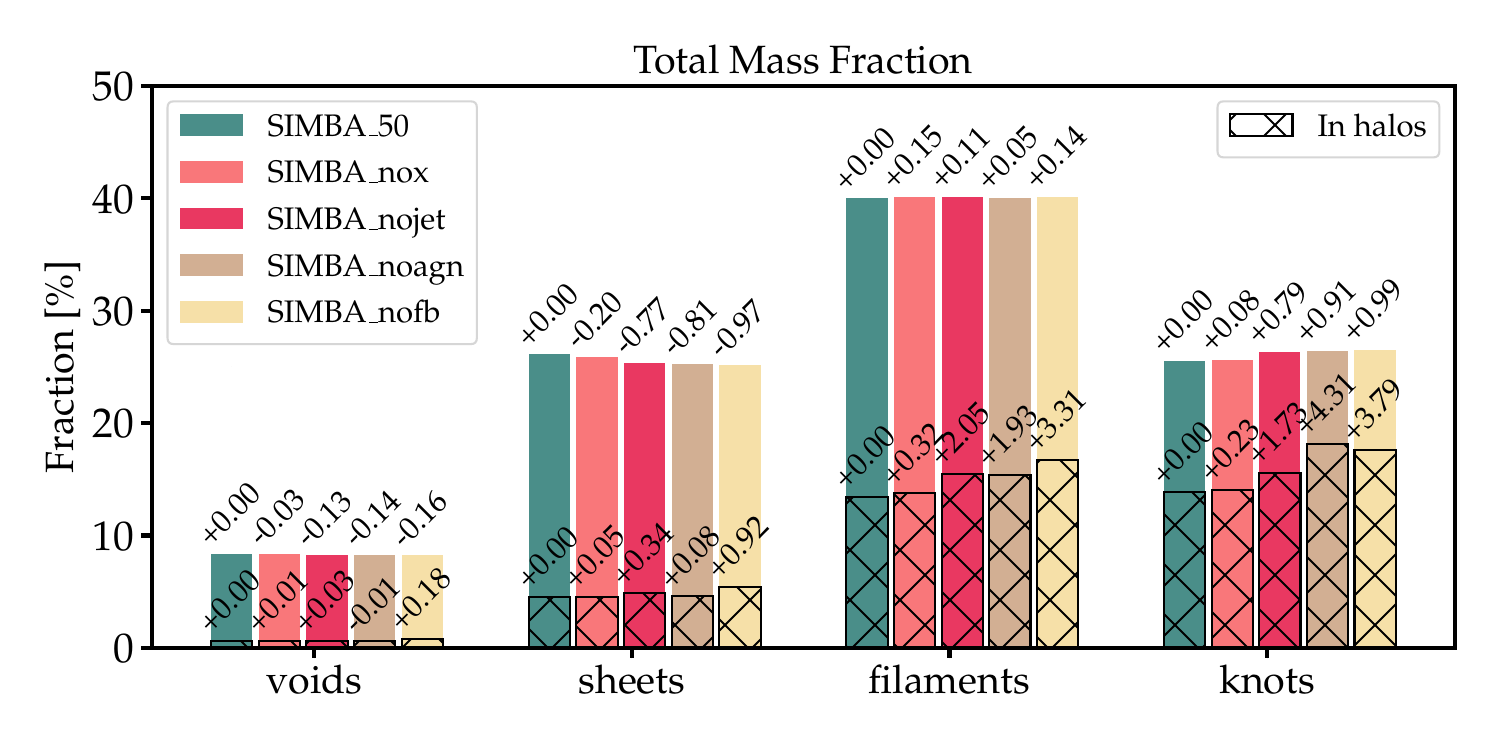}}
    \end{subfigure}
    \hspace{0.5cm}
    % \hfill
    \begin{subfigure}[b]{0.45\textwidth}
        \scalebox{0.35}{\includegraphics{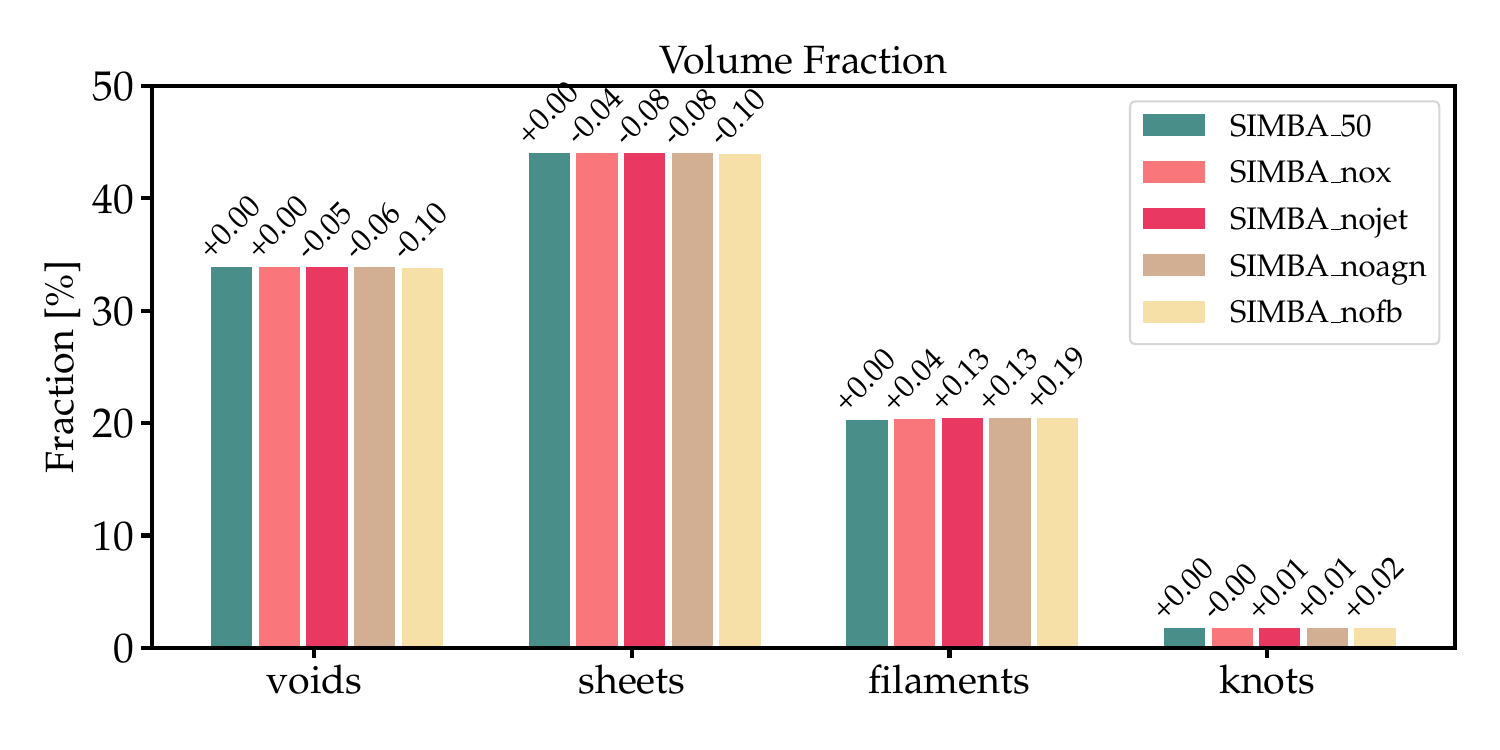}}
    \end{subfigure}

    \vspace{0.0cm}

    \begin{subfigure}[b]{0.45\textwidth}
        \scalebox{0.35}{\includegraphics{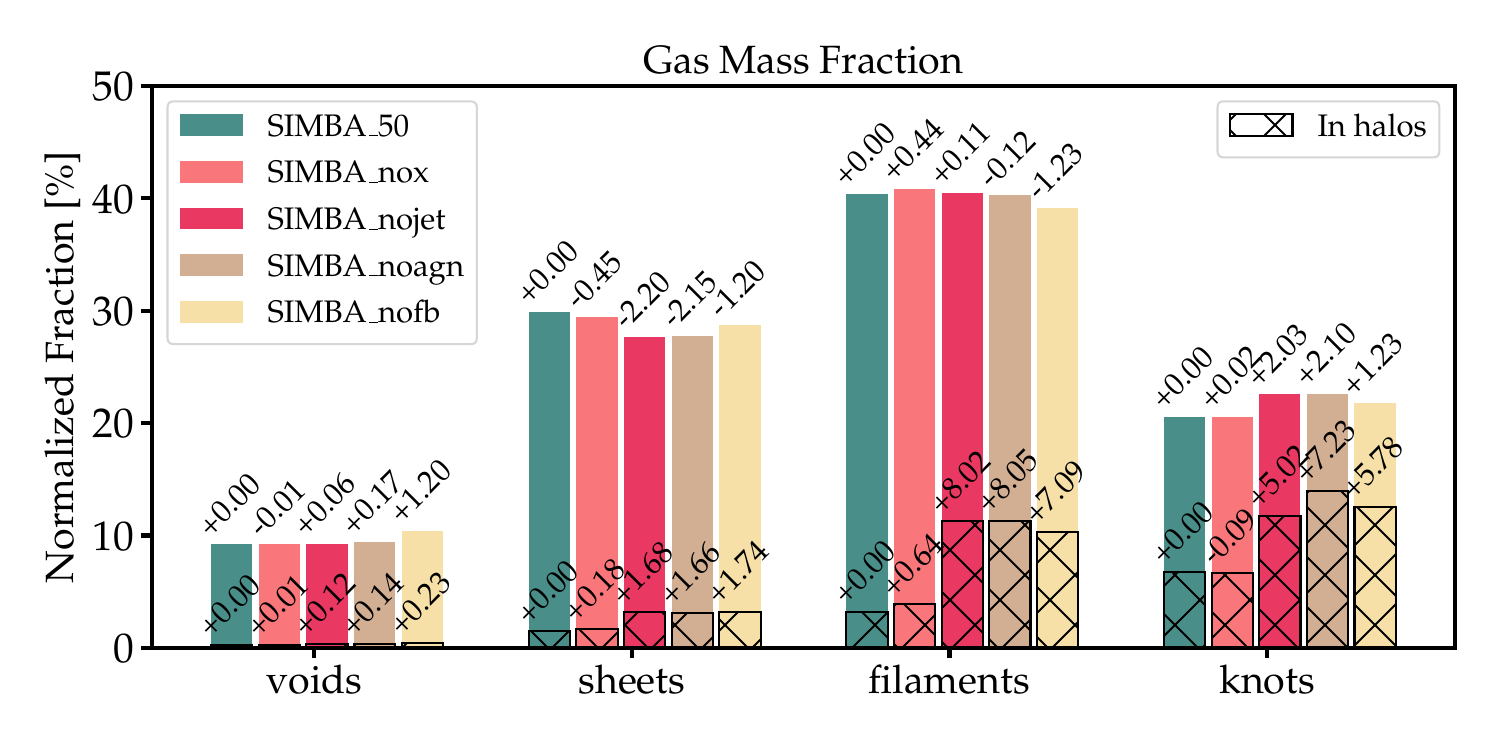}}
    \end{subfigure}
    \hspace{0.5cm}
    % \hfill
    \begin{subfigure}[b]{0.45\textwidth}
        \scalebox{0.35}{\includegraphics{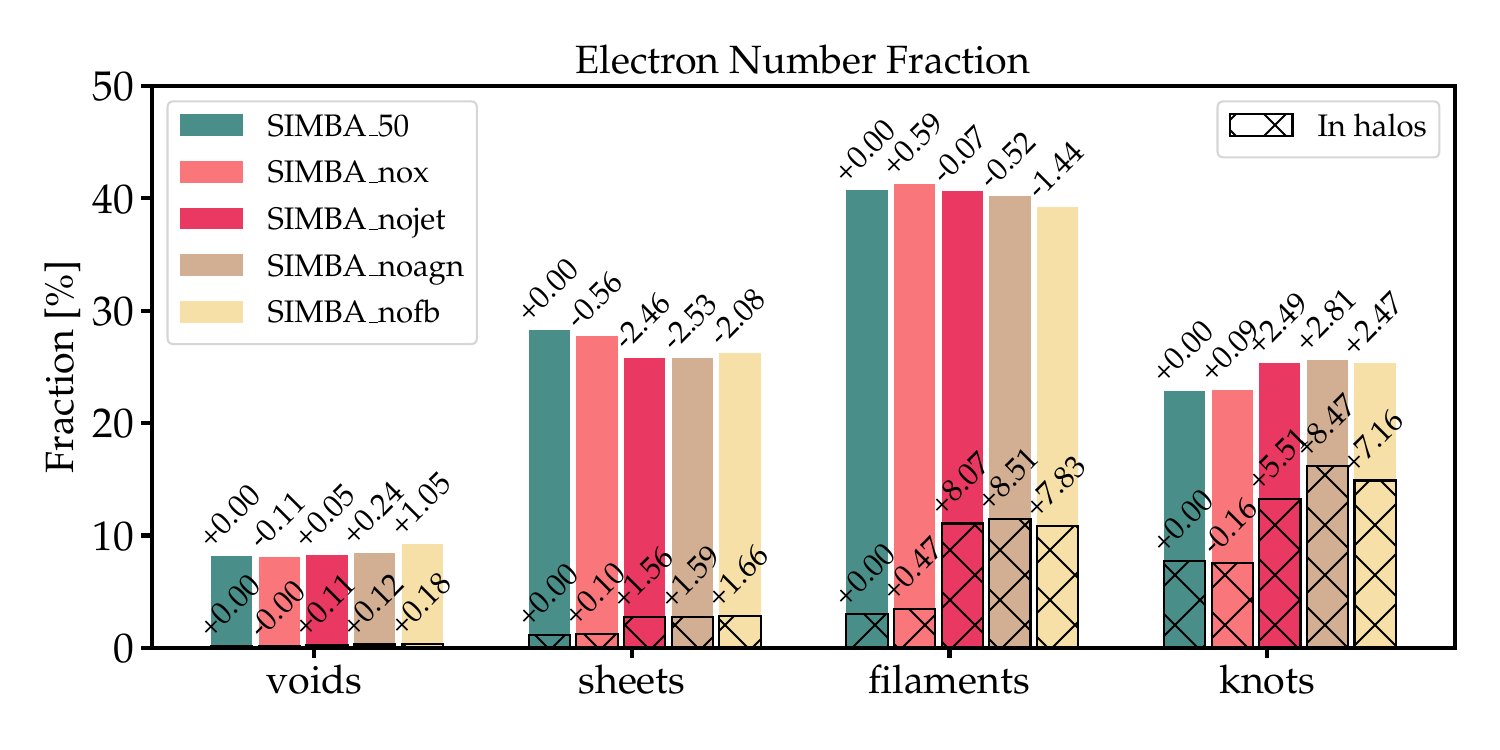}}
    \end{subfigure}
    \caption{Fractions of total matter (dark matter and baryons, upper left), volume (upper right), gas (lower left) and free electrons (lower right) that belongs to the different large-scale structures at $z=0$. The variation of feedback prescriptions (see Table~\ref{table:data_spec}) is color-coded. 
    The fraction of matter inside haloes is shown as hatched regions, except for the volume fractions. The numbers (in percentages) at the top of each bar plot indicate the relative variation of the fraction percentages with respect to the SIMBA-50 simulation.
    }\label{fig:four_in_one}
\end{figure*}

\subsection{Total mass fraction and volume fraction in the structures}

The upper left panel of Figure~\ref{fig:four_in_one} illustrates the total mass fraction, including both dark matter and baryons, within the cosmic web structures at $z=0$. 
Approximately 10\% of the mass is in voids, 25\% in sheets, 40\% in filaments, and 25\% in knots.  
Among the various feedback mechanisms, only jet feedback slightly alters the distribution of mass across these structures. 
Activating jet feedback slightly decreases the mass in knots by a fraction of 0.7\% while adding to the mass in sheets by a fraction of 0.6\%.
The results suggest the T-web classification with our choice of hyperparameters is robust against the variation of feedback models. 
Such robustness is anticipated since our classification relies on the total density field that  is predominantly shaped by dark matter without feedback influences. Besides, the influence of feedback physics on baryons is likely smoothed out under the smoothing scale of 2 cMpc/h.

The panel also illustrates the fraction of matter occupied by haloes within each component of the cosmic web (hatched histograms).  
Across all feedback variants, we observe that a growing fraction of matter is contained within haloes as the structures become denser: approximately 40\% of the matter in filaments is located in haloes, whereas in knots, this proportion increases to more than 50\%. 
Feedback physics has a greater impact on the fraction of matter in haloes, especially in filaments. 
In general, feedback prescriptions can affect halo matter fractions by a few percent. Stronger feedback results in reduced matter within haloes, as found in \citet{Khrykin:2024}. This effect is primarily observed in filaments and knots.
We interpret this as a result of stronger SMBH activities and enhanced star formation in those overdense regions. 
AGN jets eject matter and prevent more efficient mass accretion in filaments and knots;
stellar feedback regulates halo accretion in filaments.

In the upper right panel of Figure~\ref{fig:four_in_one}, we show the volume fraction of cosmic web structures in the \simba's feedback variants. 
We find that the volume fraction is highly insensitive to the sub-grid models: In all the feedback variants, the fractions of voids, sheets, filaments, and knots are 34\%, 44\%, 20\%, and 2\%,  respectively. 
The differences between the feedback variants are within 0.1\%. 
The insensitivity also arises from using the total mass density for cosmic web classification, which is nearly consistent across feedback variants.

\subsection{Gas and free electron fraction in the structures} \label{subsec:baryon_in_structrues}

We further examine the distribution of baryonic matter and free electrons within the large-scale structure. 
The lower left panel of Figure~\ref{fig:four_in_one} illustrates how gas is distributed across various parts of the cosmic web, depending on feedback. 
The observed trend is akin to that in the total mass fraction, but with a more pronounced effect from the feedback models: turning on jet feedback increases the gas fraction in sheets and decreases the fraction in knots by 2\%. 
Feedback also has a subtle impact on filaments, while their gas fraction remains almost the same.
Notably, removing stellar feedback increases the fraction of gas in voids and sheets while it decreases it in filaments and knots by 1\%.

This can be explained by recalling that the gas fraction is mainly determined by two factors: star formation and baryon expulsion.
The lack of feedback in the \nf{} run leads to more active star formation, resulting in a reduced quantity of extragalactic gas.
On the other hand, AGN feedback provides enough energy to the surrounding IGM to suppress excessive star formation, but such feedback also removes the baryons from knots. 

In haloes, we note a significant reduction in gas fraction due to jet feedback \citep[as previously seen in][]{Appleby2021,Sorini2022}. This reduction is the most significant in filament haloes, with as much as a 10\% decrease. 
AGN thermal feedback and stellar feedback also contribute to shaping the gas partition in knots, whereas X-ray heating feedback is not as efficient.
As for knots, we observe that turning off stellar feedback leads to a decrease in the gas mass fractions in haloes due to intense star formation.

The lower right panel of Figure~\ref{fig:four_in_one} illustrates the quantity of free electrons present within the structures. 
The partition of free electrons resembles the gas mass partition: jet feedback raises the free electron fraction in sheets and decreases it in knots; within haloes, jet feedback reduces the free electron fraction in knots by half, similar to the situation of the gas mass fraction.

\subsection{Baryon and IGM Gas Partition in different environments} \label{ssec:matter_part}

\begin{figure*}
    \centering
    \scalebox{0.5}{\includegraphics{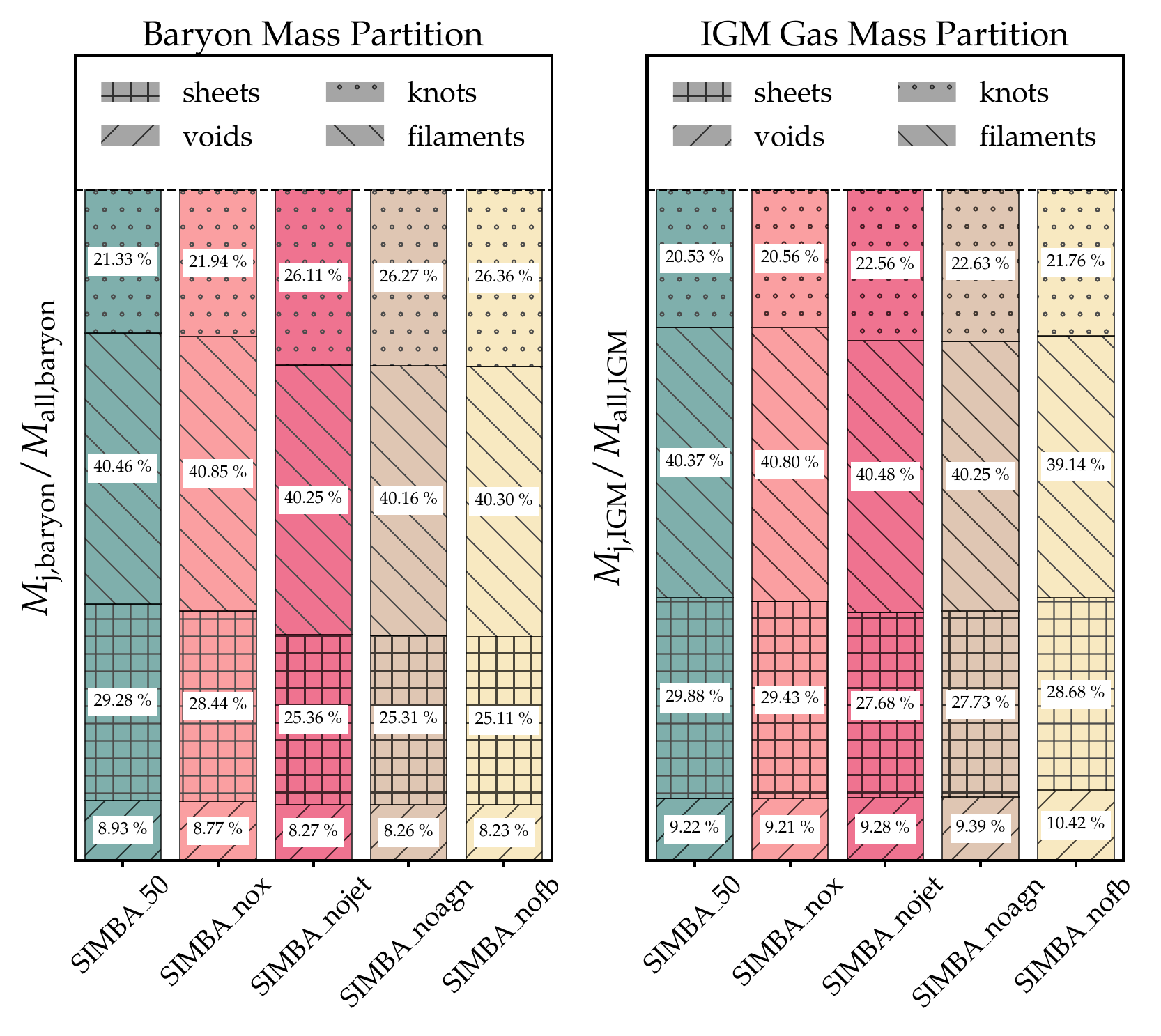}}
    \caption{
\textbf{Left}: the partition of baryon (gas, stars, and black holes) in various structures for the different feedback models. 
The fractions are normalized with the baryon mass in the whole simulation box.
\textbf{Right}: the partition of IGM gas (excludes haloes). 
The values are normalized by the total IGM gas mass in the simulation box.}
    \label{fig:fraction_normalize_by_IGM}
\end{figure*}

One goal of this study is to test whether cosmic web information helps improve the efficiency of FRB foreground mapping.

The rationale of FRB foreground mapping analysis is introduced in Appendix~\ref{appendix:FLIMFLAM_intro}. 
FRBs probe the distribution of baryon following a line-of-sight integration, which can be decomposed into distinct contributions from different cosmic components (Equation \ref{eq:DM_FLIMFLAM}).
A larger divergence of the cosmic web partition of baryon and IGM gas among the feedback variants implies that the gas partition among the cosmic web is more sensitive to the feedback prescription (Equations \ref{eq:flimflam_assumption} and \ref{eq:n_igm}). 
Consequently, such differences should be more readily detectable through FRB probes(Equation~\ref{eq:new_decomp}) when parameterized with structure-dependent baryon partitions (Equation~\ref{eq:norm}; derived in Equation~\ref{eq:f_igm_breakdown}).

In Figure~\ref{fig:fraction_normalize_by_IGM}, we present the cosmic web partition of baryon $F_{\mathrm{j,baryon}}$ and IGM gas $F_{\mathrm{j, IGM}}$ with respect to different feedback physics. 
The calculation of $F_{\mathrm{j, IGM}}$ focuses only on the IGM gas; i.e., the gas associated with haloes is excluded.
Although the effect of the AGN jet feedback is very significant on haloes, this is not automatically true for the relative matter distribution across the IGM, i.e., the distribution among voids, sheets, filaments, and knots. 
Indeed, there is only a lower than one percent effect for the baryon partition in filaments and voids from jet feedback. 
In contrast, knots and sheets show a few percent change, which remains minor compared to the impact of the jets on haloes. 
Other variations in feedback models similarly exert a negligible influence on the partitioning of IGM gas, producing differences that are less than 2\%.

Our findings indicate that various feedback mechanisms account for merely slight differences ($1\%-3\%$) in different parts of the cosmic web, normalised to either the baryon mass partition or the IGM gas mass partition. Although the smoothing scale of $2 \hcMpc$ could also contribute to diminishing the differences, it nonetheless implies that enhancing the data model, as outlined in Equation~\ref{eq:new_decomp}, does not significantly improve the results at the current resolution of the density reconstruction.

\section{The local dependence of IGM gas fraction on environment} \label{sec:f_gas}

In Section \ref{sec:matter_partition}, we have analysed the distribution of IGM gas through the global partition among large-scale web structures.
Extending the definition of $\fgas$ in \citet{Khrykin:2024},
 we now focus on the gas fraction in individual voxels with the goal of understanding the distribution of IGM gas inside each type of structure.

Following the definition of $\fgas$ in Equation~\ref{eq:f_gas}, we consider the correlation of the IGM gas distribution relative to the cosmic web in Section~\ref{subsec:corr_f_gas_cw}, then provide a quantitative description in Section~\ref{subsec:f_gas_ovd}.
In Section~\ref{subsec:dm_test}, we test the impact of the correlation on the parameter estimation in FRB foreground mapping. 

\begin{figure*}
    \centering
    \scalebox{0.4}{\includegraphics{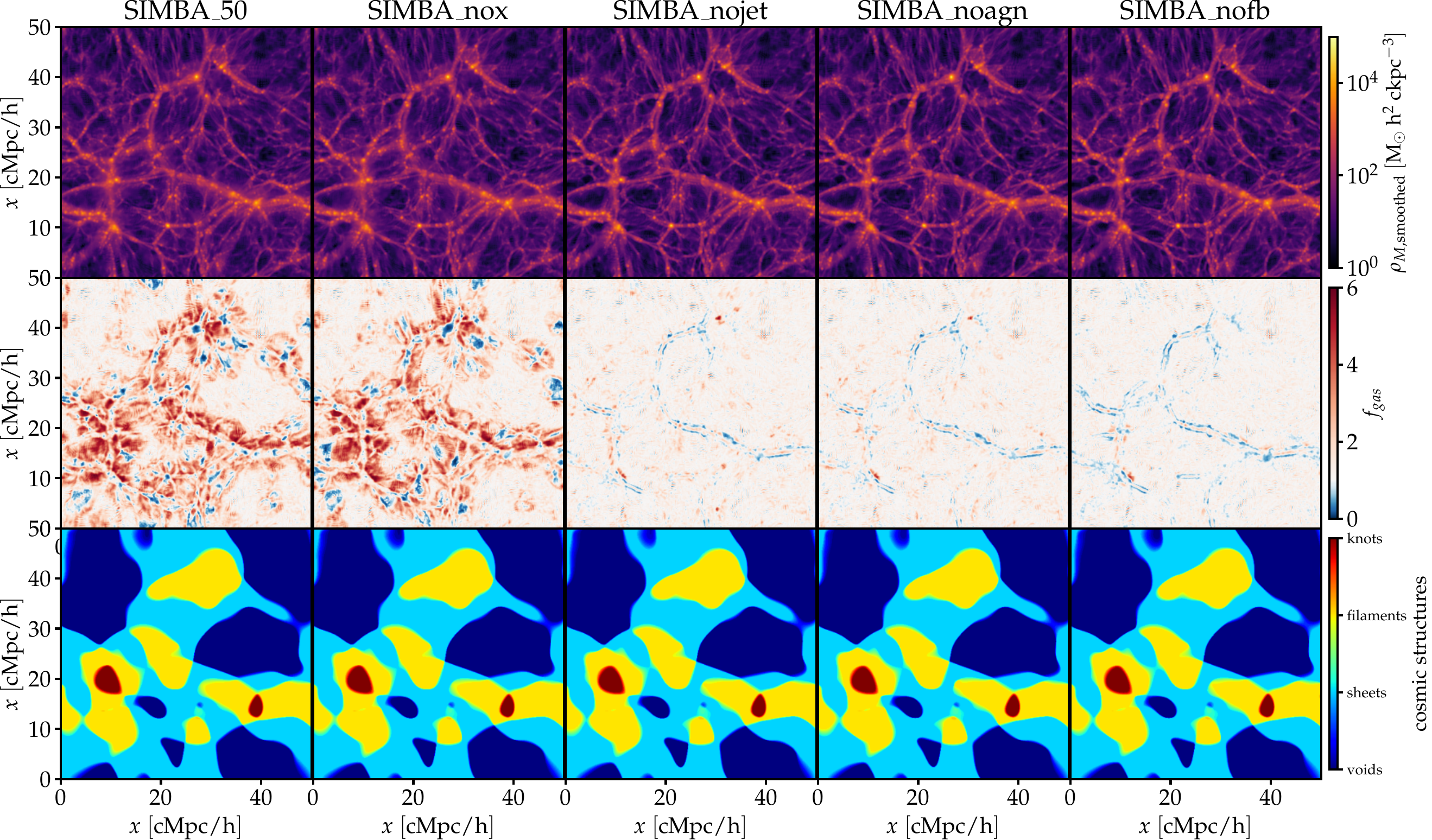}}
    \caption{\textbf{Left to right}: Slices (1 cMpc/h thickness) of \simba feedback variants ({Simba}-50, {Simba}-nox, {Simba}-nojet, {Simba}-noagn, {Simba}-nofb) showing the density field, $\fgas$ value and associated cosmic web classifications. 
    \textbf{Top row}: The slice of density field taken from the $z=0$ snapshot. 
    \textbf{Middle row}: The $\fgas$ distribution in the slices. Note the colormap is chosen to indicate relative gas deficits ($\fgas<1$) as blue and gas excess ($\fgas>1$) as red. 
    \textbf{Bottom row}: Corresponding averaged classification field, performed with $\lambda_{th} = 0.16$. 
    The calculation of $\fgas$ is based on $100$ ckpc/h Gaussian smoothed field in order to suppress the numerical artifacts.
    The deficit and excess of $\fgas$ is spatially correlated with the cosmic web classification. 
    }
    \label{fig:f_gas_vis}
\end{figure*}

\begin{figure*}
    \centering
    \scalebox{0.4}{
    \includegraphics{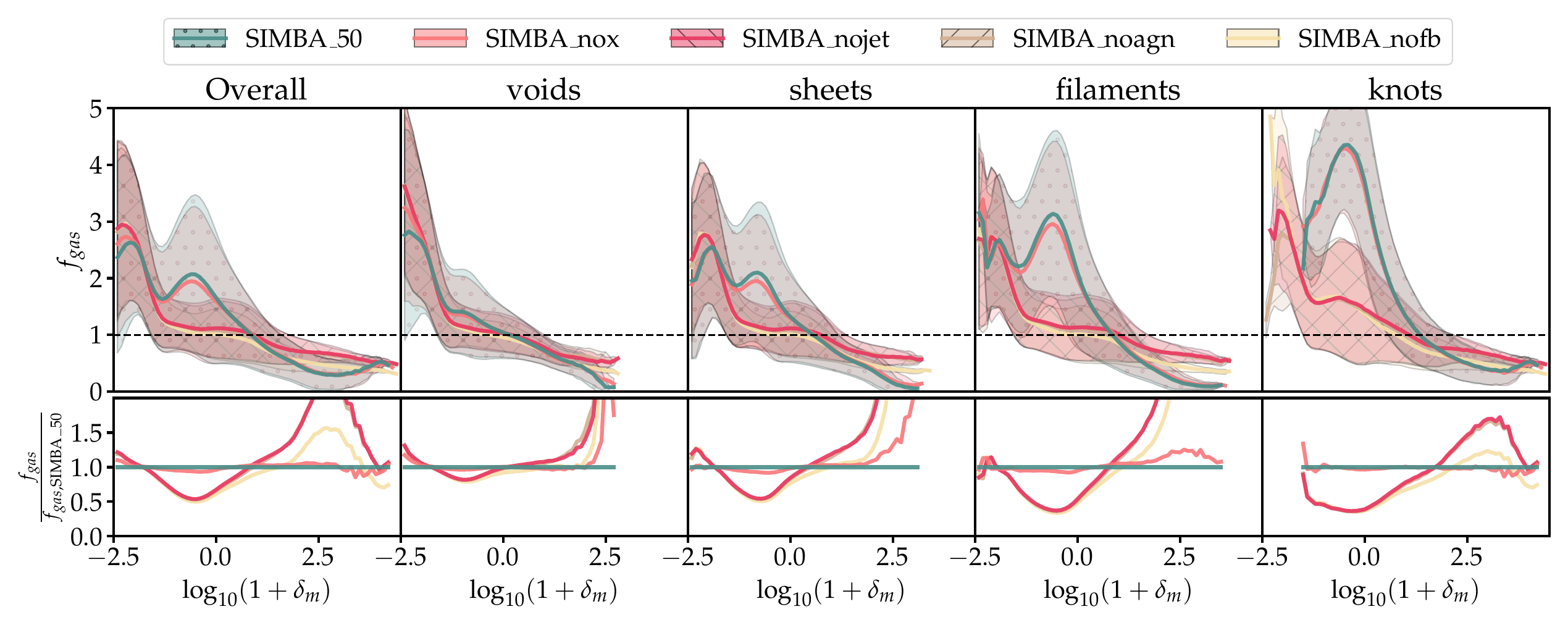}}
    \caption{\textbf{Upper panels:} the gas fraction $\fgas$ as a function of overdensity $(1+\delta_m)$ under $0.1$ cMpc/h Gaussian smoothing, in every T-web structure and for all \simba feedback variants. 
    \textbf{Lower panels:} the ratios of $\fgas$ values between different feedback variants and \ff{} run.
    \textbf{Left to right}: the relation in the whole simulation volume, in voids, in sheets, in filaments and in knots.
    The shaded regions represents $1\sigma$ range of the $\fgas$ distribution. 
    Within the \nx{} and \ff{} runs, we observe a U-turn pattern where $\fgas < 1$ (gas deficit) linked to haloes, whereas in the diffused IGM, it appears that $\fgas\sim 2$.
    The amplitude of the bump is stronger when the simulation volume is separated according to the cosmic web classification: the fraction reaches 
    $\fgas>3$ in filaments and knots.
    The hatched regions represent the 1-$\sigma$ range of $\fgas$ at a given overdensity for each feedback variant.}
    \label{fig:f_gas}
\end{figure*}

\subsection{The correlation between $\fgas$ and the cosmic web}\label{subsec:corr_f_gas_cw}

\begin{figure*}
    \centering
    \scalebox{0.4}{
    \includegraphics{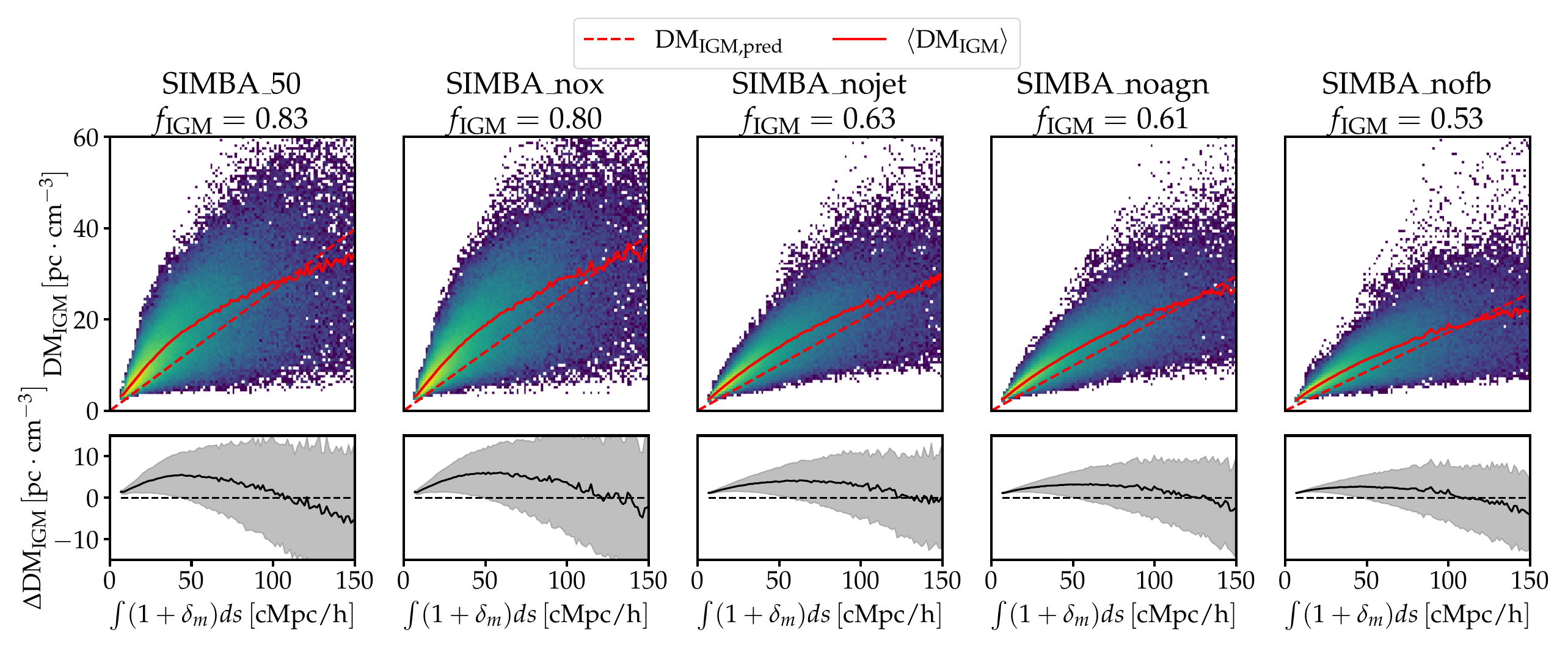}}
    \caption{The relation between the overdensity integral \intdelm{} and the measured $\mathrm{DM}_\IGM$ in \simba's feedback variants at $z=0$. 
    In the upper panels, 2D histograms of the two integrated quantities are plotted for the five feedback variants. 
    The solid red lines show the mean $\langle \mathrm{DM}_\IGM\rangle$ for a given \intdelm{} bin, measured in the simulation; the red dashed lines follow the prediction of $\mathrm{DM}_{\mathrm{IGM, pred}}$ in Equation~\ref{eq:flimflam_assumption}.
    The baryon fraction in IGM $\figm$ adopted for $\mathrm{DM}_{\mathrm{IGM, pred}}$ calculation is listed on the top of each panel.  
    As a reference, FLIMFLAM reported $\figm = 0.59_{-0.10}^{+0.11}$ with 8 FRB sightline in the first data release.
    There is a notable excess of measured $\langle \mathrm{DM}_\IGM\rangle$ for $\intdelm{}<100\,\mathrm{cMpc/h}$, especially in simulations with strong feedback. 
    In the lower panels, the excess $\Delta \mathrm{DM}_\IGM = \langle \mathrm{DM}_\IGM\rangle- \mathrm{DM}_{\mathrm{IGM, pred}}$ is quantified as black solid lines; the gray shadow regions represent a 1$\sigma$ error of the measured $\mathrm{DM}_\IGM$ distribution. 
    }
    \label{fig:DM_test}
\end{figure*}

In Figure~\ref{fig:f_gas_vis}, we show the $\fgas$ distribution for a $1$ cMpc/h slice of \simba feedback variants (second row) together with the density field (first row) and the classification of the cosmic web (third row).

For \ff{} and \nx{}, an excess of $\fgas$ is found at the outskirts of filaments and knots, while the core regions of knots show a deficit due to jet feedback. 
For the other feedback variants, the variation of $\fgas$ is mainly present in filaments; a coaxial structure of gas deficit (surrounding) and gas excess (center) forms along the visual filaments, i.e., the filaments that are clearly visually distinguishable in the matter distribution.
It is worth noticing that the `visual' filaments are not necessarily equivalent to the T-web filament defined in Section \ref{subsec:class}, mostly because of the coarse resolution scale adopted in the T-web reconstruction.
The detection of the `visual' filaments requires the application of filament finders using galaxy positions as tracers of the density field, such as DisPerSE \citep{Sousbie:2011a, Sousbie:2011b}, T-Rex \citep{Bonnaire_2020}, or 1-DREAM \citep{Awad:2023}, which is beyond the scope of this work, which requires a more volumetric method in order to unambiguously assess the IGM baryon partition.

Figure~\ref{fig:f_gas_vis} shows that multiple types of physics at different scales contribute to shaping the spatial correlation. 
In the absence of jet feedback, the primary mechanism that influences the gas distribution is accretion. Accretion results in the aggregation of gas within high-density regions, consequently leading to a gas deficit at the outskirts of `visual' filaments. 
In contrast, jet feedback drastically changes the correlation by ejecting gas into the IGM, creating a gas deficit within haloes and a surplus in the surrounding filaments. 

These findings bring a higher resolution perspective into our results of
Section \ref{sec:matter_partition}, where we found minimal changes in the IGM gas partition among cosmic web environments. Our findings indicate that jet feedback significantly alters the distribution within the cosmic web structures. 
However, the impact is limited to regions near massive structures and does not extend beyond the cosmic web size (a few to ten Mpc) defined in Section \ref{subsec:class}.

\subsection{The dependence of $\fgas$ on the overdensity}\label{subsec:f_gas_ovd}

Inspired by \citet{Sorini2022} and \citet{ Khrykin:2024} reporting that the gas fraction $\fgas$ within haloes depends on the halo mass for different feedback variants, we investigate the dependence of $\fgas$ on the local overdensity $\delta_m = \rho_m/\bar{\rho}_m-1$. 
We show the dependence of $\fgas$ on $(1+\delta_m)$ in the leftmost panel of Figure~\ref{fig:f_gas}.
We find a similar trend to Figure 3 of \citet{Khrykin:2024}. 
In the following, we describe the trends from high to low over-density values.
For the volume with $(1+\delta_m)\gtrsim100$ where haloes usually reside, the $\fgas$ for \nx{} and \ff{} shows a U-turn shape: it goes down until $\log(1+\delta_m ) \sim 2.5$, then rises again in the more massive regime. 
For relatively lower overdensities associated with diffuse IGM, $0.1<(1+\delta_m)<10$, we find a bump 
with a peak $\fgas\sim2$ at $(1+\delta_m)\sim 0.3$ for these two prescriptions, due to jet feedback that ejects halo gas to the IGM and prevents the accretion of IGM gas. 
As a result, the gas decouples from dark matter and remains in the diffuse IGM, leading to an excess of $\fgas$ at $0.1<(1+\delta_m)<10$. 
The other three feedback models, \nf{}, \nagn{}, and \nj, show a steadily decreasing relation between $\fgas$ and $(1+\delta_m)$ for high overdensity, but their $\fgas$ values are always higher than those of \nx{} and \ff{} for such large overdensities. 
For diffuse gas in moderate nonlinearities ($0.1 < (1+\delta_m) < 10$) and no jet feedback, assuming the gas fraction to be at the cosmic value $\fgas \approx 1$ provides a nearly unbiased approximation;
For extremely low overdensities $1+\delta_m < 0.01$, we find that all prescriptions have $\fgas > 1$, which is likely due to the Gaussian smoothing operation applied to the density fields. 
These regions are more easily affected by the smoothing, as their intrinsic density is small compared with the change introduced by the smoothing. 
Therefore, we argue that those $\fgas$ values do not have a physical origin. 

When taking an environment-based view, we find that the U-turn trend of $\fgas$ related to jet feedback is only present in cosmic web knots; for the other cosmic web structures, $\fgas$ can be as low as zero at high overdensities.
Notably, distinguishing the volume based on the cosmic web structures isolates the signal of $\fgas\neq1$ in the diffuse IGM.
In knots and filaments, the peak of $\fgas>3$ indicates that the mean baryon content in the diffuse IGM can be three times higher than the expected value, assuming gas traces the density field exactly. 
Within cosmic voids, however, the excess almost disappears.

We interpret the U-turn at $(1+\delta_m)>100$ and the bump at $0.1<(1+\delta_m)<10$ as a result of strong jet feedback in the \nx{} and \ff{} runs. 
For one thing, the jet facilitates the redistribution of baryons between the IGM and haloes, with a pronounced effect within intermediate-mass haloes. 
In these systems, the activity of SMBHs is better fuelled compared with smaller haloes, and the ejected gas has a higher probability of escaping the gravitational potential well compared to that in larger haloes.
Thus, the bump indicates the destination of gas expelled from haloes --- they reside in the diffuse IGM, leading to the gas fraction $\fgas>1$. 
In the absence of jet feedback, it is star formation that dominates the variation of $\fgas$ in haloes. 
This results in the monotonically declining trend of $\fgas$ observed in \nj{}, \nagn{}, and \ff{}. 
Both stellar feedback and AGN thermal feedback still play a role in controlling star formation, which explains why the $\fgas$ value is marginally higher in the \nj{} scenario. 
Without jets, short-range galaxy feedback in \simba has a minimal impact on baryon content, which leads to $\fgas$ being almost one.
The trend becomes stronger because the cosmic web classification filters the mass of the associated haloes, e.g. massive galaxy clusters preferentially reside in cosmic web knots.

% \subsection{A test of baryon excess on $\mathrm{DM}_{\mathrm{IGM}}$} \label{subsec:dm_test}
\subsection{Deviation in predicted $\mathrm{DM}_{\mathrm{IGM}}$ arising from feedback-induced baryon excess} \label{subsec:dm_test}

% Following the discovery in Section \ref{subsec:f_gas_ovd}, we further examine the impact of baryon excess in the context of the FRB dispersion measure. 
Following the results in Section \ref{subsec:f_gas_ovd}, we realised that the $\fgas$ dependence on matter density is an overlooked factor when tracing the baryons in the intergalactic medium with FRBs.
Below, we perform a first analysis of the impact of baryon excess in the context of the FRB dispersion measure.
We utilised only the snapshot $z=0$ for this test due to the limitations of the redshift in our study.
We leave to future work a comprehensive analysis that goes beyond the qualitative trends presented in this section. Future work should include the construction of mock lightcones and the modelling of survey systematics.

Following the formulation detailed in $\mathrm{DM}_\mathrm{IGM}$ within FLIMFLAM (see Equations \ref{eq:n_igm} in Appendix A), we first integrate $\mathrm{DM}_\mathrm{IGM} = \int n_{e, \IGM} ds$ and \intdelm{} across the 50-cMpc/h side boxes. 
Integration is performed for $500\times500$ gridded FRB lines of sight along the $x$ direction with an even spacing of $0.1\,\mathrm{Mpc/h}$, using the electron number density field and the matter density field with $R=0.1\,\mathrm{cMpc/h}$ Gaussian smoothing. 
The details of the FLIMFLAM model are introduced in Appendix \ref{appendix:FLIMFLAM_intro}.
For clarity, we briefly summarise the assumptions entering the DM modelling: we assume a fully ionised IGM and neglect small-scale host contributions (Equation \ref{eq:flimflam_assumption}).
Since we work only on the snapshots at $z=0$, we decide to omit the scale factor $1/(1+z)$. Indeed, the effect of lightcone integration, implemented by placing an imaginary observer in a corner of the simulated box, would be negligible (sub-percent) because of the small box size $50$ cMpc/h.
We calculated the prediction $\mathrm{DM}_\mathrm{IGM, \mathrm{pred}}$ from the integration \intdelm{} of the FRB sightlines (Equation~\ref{eq:flimflam_assumption}), and compared it with $\mathrm{DM}_\mathrm{IGM}$ measured in the simulation. 
In FLIMFLAM, the predicted $\mathrm{DM}_\mathrm{IGM, \mathrm{pred}}$ is considered to be proportional to \intdelm{}, up to a factor (Equation \ref{eq:n_igm}) related to the fraction of baryons residing in the IGM, $\figm$, for which we adopted the value measured in SIMBA's feedback variants at $z=0$. 
The results are shown in Figure \ref{fig:DM_test}.

Regarding the sightlines that do not intersect any massive structures ($\intdelm<100\,\mathrm{Mpc/h}$), the mean value $\langle\mathrm{DM}_\IGM\rangle$ exceeds the predictions for all variations in feedback. 
On average ($\intdelm = 50 \,\mathrm{cMpc/h}$), deviations can reach about $5 \,\mathrm{pc\cdot cm^{-3}}$ within both \ff{} and \nx{} simulations. 
As the feedback effect diminishes, Equation~\ref{eq:flimflam_assumption} yields a better prediction of $\mathrm{DM}_\IGM$.
For $\intdelm > 100 \,\mathrm{cMpc/h}$, the observed DM falls below the predicted value. 
This occurs as the sightline passes through massive structures, where star formation depletes the gas, or the feedback mechanism drives the gas away. 
These findings are in agreement with Figures \ref{fig:f_gas_vis} and \ref{fig:f_gas}, where we report an excess in $\fgas$ for $0.1<1+\delta_m<10$ when the AGN jet feedback is present; at higher overdensities ($1+\delta_m>100$), all feedback variants have a $\fgas$ dropping below $1$.
Interestingly, the underestimate of $\mathrm{DM}_\mathrm{IGM}$ also occurs in the \nj{} and \nagn{} runs, though it is weaker than the simulations with jet feedback. 
This underestimate hints at the effect of AGN thermal feedback or stellar feedback on the removal of gas from haloes, as implied in the ``filaments'' panel in Figure \ref{fig:f_gas}.
Without the presence of any feedback (\nf{} run), the assumption of a global $f_{\IGM}$ becomes a more reliable approximation. %solid approximation. 
In summary, our test on $z=0$ snapshots indicates a growing risk of underestimating the dispersion measure when the stronger impact feedback on the IGM is not properly modelled through a density-dependent $\fgas$. 
From the perspective of DM dispersion measure decomposition, the underestimation will introduce a bias in the inference of $\figm$. 
Dedicated work on simulation data across $0<z<1$ is essential to accurately assess the impact and integrate it into the FRB DM data model.

\section{Discussion} \label{sec:discussion}
In this work, we adopted the T-web method to detect the cosmic web at observationally motivated scales in \simba simulations with different feedback variants.
Based on the classification, we investigate the impact of feedback on modifying the relative distribution of matter and gas across the cosmic web. 

\subsection{Comparison with previous works}

In this section, we compare our results with previous works from the simulation and observation communities.
In Section~\ref{subsec:baryon_in_structrues}, we find that jet feedback has a limited effect on the gas partition across cosmic web environments (defined at observationally motivated scales). 
However, when taking a smaller-scale view, we find that the feedback of the AGN jet influences haloes, particularly in filaments (upper left panel of Figure~\ref{fig:four_in_one}), largely due to baryon removal from haloes (lower left panel of Figure~\ref{fig:four_in_one}). 
In turn, the IGM gas around massive structures (i.e., haloes) tends to exhibit a gas excess as jet feedback moves halo gas to the IGM (Figure \ref{fig:f_gas_vis} and \ref{fig:f_gas}). 

The dependence on feedback for baryon partition between the IGM and halos has been widely studied from a simulation-orientated aspect.
\cite{Sorini2022} found that in \simba, AGN jets effectively remove baryonic matter from $M_h>10^{12} M_\odot$ haloes at redshift $z<2$. 
Additionally, using the same set of simulations, \cite{Khrykin:2024} reported that AGN jets reduce the CGM mass fraction by a factor of six at $z=0.1$ in haloes with $10^{12}M_{\odot}<M_h<10^{14}M_{\odot}$. 
Our study agrees with this trend in the baryonic fraction distribution and $\fgas$ for $(1+\delta_m)>100$. 

Additionally, \cite{Sorini2022} studied the distribution of different baryonic phases within haloes in the mass range $10^{11} - 10^{14} \, \rm M_{\odot}$ in the \simba{} suite of simulations. Noting that, except for the no-feedback variant, haloes are generally baryon deficient, they showed that the halo-centric distance enclosing 90\% of the cosmic baryon mass fraction varies considerably with halo mass and redshift. 
The scale was found to be sensitive to the feedback strength and can thus be used to constrain the feedback model. 
These results were independently confirmed by \cite{Ayromlou:2023}, who also extended the study of the dependence of the aforementioned length scale (adopting a slightly different operational definition) to the IllustrisTNG and EAGLE simulations as well.
Although the recent works mentioned above imply excess baryons around haloes, as seen in Figure~\ref{fig:f_gas_vis}, they emphasised a halo-focused perspective motivated by X-ray and SZ observations \citep[see also][]{Angelinelli:2022, Angelinelli:2023}, in contrast with our study's motivation from the FRB point of view, in which the dispersion measure is a volume-weighted probe of cosmic gas.

Although the baryon deficit in haloes implies an enhanced baryon fraction in the vicinity of haloes, the influence of the latter is largely overlooked in much of the literature.
A rare perspective from the IGM side is provided by \citet{Wang:2024}, where the authors observed a baryon excess in low overdensity environments using simulation suites including \simba, Illustris, IllustrisTNG, and EAGLE.
Interestingly, according to \citet{Wang:2024}, such excess exists in all simulation suites: in \simba and Illustris, the excess is overall stronger and overdensity dependent; In IllustrisTNG and EAGLE, the excess shows a milder dependence on overdensity and is overall weaker.
These results are in broad agreement with our findings in Figures \ref{fig:f_gas} and \ref{fig:DM_test}. In \ff{} and \nx{} runs, jet feedback causes significant $\fgas$ excess and subsequently an underestimation of FRB DM; the excess is less notable in runs without jet feedback, but it will still introduce bias to $\mathrm{DM}_\mathrm{IGM}$ estimation. 

% DANIELA has comments here
Prior simulation studies motivated by absorption line observables, such as the work by \citet{Oppenheimer:2012}, report a significant impact of feedback physics on the thermal and chemical evolution of the WHIM, as seen in the equivalent width distributions of metal lines. 
Also, \citet{Martizzi_2019} shows that the WHIM resides mainly in filaments. 
Although they do not compare different feedback models, feedback is expected to shape the baryon distribution across the cosmic web, given the link between gas phase and partition.
However, in Figure \ref{fig:fraction_normalize_by_IGM}, we only observed a marginal difference in the IGM gas mass partition among the feedback variants.
The difference in results shows that the large-scale gas distribution beyond halos is only weakly linked to metal enrichment or gas phase. 
This agrees with \citet{Suresh:2015}, who reports nearly constant baryon content within $3 R_{\mathrm{vir}}$ of halos, despite large variations in gas phase fractions and CGM metallicity across different feedback models.
Conversely, at smaller scales, the gas distribution can be sensitive to the feedback (Figure \ref{fig:f_gas}) and is possibly detected via FRB observations. 

Among FRB-motivated simulation works, \cite{Walker:2024} aligns closely with the objectives of our study. 
Their research on the TNG-300 simulation categorises cosmic structures into haloes, filaments, and voids with density-based criteria and assesses their influence on FRB dispersion measures. 
In their findings, the filament baryon fraction decreases from approximately $90\%$ at $z=5$ to around $65\%$ at $z=0$, where the latter closely matches our ``sheets+filaments'' baryon fraction. 
They also found lower DM variance for FRBs sightlines in voids and suggested utilising this for selecting FRBs that mainly pass through low-density structures to probe cosmological parameters. 
Our findings, illustrated in Figure~\ref{fig:f_gas}, endorse this approach: voids have the least feedback model dependence, which simplifies the interpretation of the DM analysis.

From the FRB observation perspective, \cite{Hussaini:2025} recently examined the link between the excess of DM of FRB, denoted as $\Delta \mathrm{DM}=\mathrm{DM}_{\mathrm{cosmo}} - \langle\mathrm{DM}_{\mathrm{cosmo}}\rangle$, and foreground large-scale structures using galaxy tracers. 
The study reported a weak but statistically significant correlation between DM excess and galaxy density, along with a clear DM excess for FRB sightlines with an impact parameter less than approximately $3\,\mathrm{Mpc}$. 
These results support our findings that feedback contributes to the IGM $\fgas$ excess, especially near the massive structures, as shown in this work.

The study conducted by \cite{Hsu:2025} also reported a positive correlation between $\mathrm{DM}_{\mathrm{cosmic}}$ excess and galaxy number density with localised FRBs and photometric galaxy catalogues. 
The authors suggested that fluctuations in ionised baryons influence the dispersion measure (DM) of FRBs on a scale of several megaparsecs. 
Although more localised FRBs and further spectroscopic follow-ups are necessary to confirm this trend, it can potentially be explained by the feedback or environmental influence of $\fgas$, as demonstrated in \ref{fig:f_gas}.
Ongoing galaxy survey programmes, such as FLIMFLAM targeted at FRB fields, or DESI \citep{DESI:2022} and the Subaru PFS Galaxy Evolution Survey \citep{Greene:2022} on wider fields, are expected to provide a substantial dataset that facilitates the characterisation of FRB sightline environments, thus playing a crucial role in validating the correlation.

To summarise, The baryon excess driven by feedback in the outskirts of massive halos and along cosmic filaments has been identified in previous simulation-based studies, although such effects are typically neglected in current FRB modelling frameworks.
This environment-dependent baryon excess is also supported by recent FRB–galaxy cross-correlation measurements.

\subsection{Choice of simulation suite and parameters}

Our analysis relies on specific choices of simulation suites and cosmic web classification parameters. 
Below, we assess how those choices could affect the generality of our conclusions.

In this study, we adopted the feedback variants of the SIMBA simulation suite to analyse the impact of feedback on the baryon distribution across the cosmic web. 
The setup of feedback variants allows us to quantify the effect of various feedback mechanisms in an incremental way. 
However, concerns may arise regarding the reliability of the jet model in the \simba simulation, considering the strong assumptions about the direction of the highly collimated jet launch. 
Furthermore, simulation data sets such as CAMELS \citep{Paco:2021CAMELS} suggest that \simba predicts the thermal evolution of the IGM gas distinctly from other cosmological simulations such as TNG and EAGLE. 
In fact, observations of the IGM favour the \simba jet feedback model in many cases.
For instance, \citet{Christiansen:2020} and \citet{tillman_efficient_2023} find that jet feedback in \simba provides better agreement with observations of the Ly$\alpha$ forest mean flux decrement and the column density distribution of absorbers, respectively, in the low redshift ($z<1$) Lyman-$\alpha$ forest. 
The jet feedback model in \simba is also the most likely explanation for the `missing' Lyman-$\alpha$ absorption around several $z\sim 2$ galaxy protoclusters, as found in \citet{Dong:2023, dong2024effect}.
On the CGM scale, the results are more mixed, and comparisons with observations are less definitive.
\cite{Dav__2019},  \cite{Robson:2020}, and \cite{Jennings:2023} showed that \simba nicely matches X-ray observations, such as the halo baryon fraction of groups, stacked X-ray luminosity vs. central galaxy K-band luminosity, and X-ray scaling relations, as well as or better than many other models. However, at higher redshifts, \cite{Sorini:2020} reports that the Lyman-$\alpha$ transmission around close quasar pairs in the \simba{} simulation does not match the observations by \cite{Prochaska:2013} within a transverse separation of $150 \, \rm kpc$.
Recently, \citet{Zhang:2025} reported that \simba predicts well the $L_X - M_*$ (X-ray luminosity versus stellar mass) relation observed by eROSITA, and remains in broad agreement with the observation on the $L_X - M_{500c}$ (X-ray luminosity versus halo mass) relation. 
In contrast, \citep{Oppenheimer:2021} noted that \simba under-predicts central hot gas densities and over-predicts inner entropy profiles in galaxy groups.
Similarly, the Gizmo-\simba runs from the 300 Cluster Project likewise show lower central densities than observed in clusters~\citep{Cui:2022}.
% DANIELA: soften or remove this!
These results suggest that SIMBA may over-evacuate the central regions of massive haloes.
Despite this, \simba's highly collimated jet model does not seem unrealistic and may, in fact, be favoured by larger-scale observations of the CGM and IGM \citep{Christiansen:2020, Appleby2021}, implying that significant gas redistribution and heating from AGN feedback is plausible.

The choice of hyperparameters can affect the classification results.
The impact of $\lth$ choice is presented in Appendix \ref{appendix:convergence_test}. 
As shown in Equation \ref{eq:classification}, $\lth$ controls the separation of structures. 
We notice that enlarging this parameter leads to thinner filaments, similar to the trend reported by \citet{Forero_Romero_2009} and \citet{Martizzi_2019}. 
Thinner filaments allow feedback to reach sheets or voids more easily, which can make cosmic-web–related diagnostics more sensitive to feedback. 
However, detecting thinner filaments is simultaneously limited by the spatial resolution achievable in observations.
The smoothing length $\sml$ filters the characteristic scale for cosmic web structures.
As demonstrated in Section~\ref{sec:cls}, the density fields of the \simba variants exhibit little visual difference, resulting in a similar cosmic web classification of the structures. 
In Section~\ref{ssec:matter_part}, we examine the impact of feedback on the matter distribution among the various cosmic web components of the IGM and report a minimal change in mass fraction, up to a few percent, which is related to our choice of observation-motivated smoothing scale, 2 cMpc/h.
Our results are in agreement with \cite{Ni:2023}, where the authors note that the effects of jet feedback in \simba gradually shrink on the scale beyond one megaparsec.

\cite{Sunseri_2023} studied the effect of baryon feedback with the TNG-300 simulation and its dark-matter-only counterpart. 
The authors deployed the NEXUS+ algorithm to define the cosmic web, which improved the T-web method by adopting the log-density field as input and using multiple smoothing kernels to capture the structures of different scales. 
The NEXUS+ algorithm visually suggests a multi-scale filament classification, compared to the fixed smoothing length in the T-web method used in this work with, yet the resulting mass fraction discrepancy for filaments is only $1\%$.
The authors found that haloes lose about 10\% of their mass due to the presence of baryon feedback.
A similar trend can be found in the gas fraction shown in Figure~\ref{fig:four_in_one}, where turning on the jet feedback causes the loss of baryons in haloes.

\subsection{Implication on FRB foreground mapping as a feedback probe}

As mentioned in Section~\ref{sec:matter_partition}, galaxy feedback of any form only causes percent-level relative changes in the IGM gas partitioning across the different components of the cosmic web at the scale of $2$ cMpc/h.
Although the partition between the halo CGM gas and broader IGM has been shown to be sensitive to galaxy feedback prescriptions \citep{Khrykin:2024},
there appears to be limited additional information from considering the gas fractions between different cosmic web components (as in Equation~\ref{eq:new_decomp}) at the scales we can actually reconstruct the web in observations, especially in current generations of the FRB survey FLIMFLAM \citep{Lee_2022, Khrykin:2024flimflam}.

% the second result
Nevertheless, it is meaningful to define an environment-dependent (or density-dependent) $\fgas$ to allow for a more precise $\mathrm{DM}_{\IGM}$ estimation. 
When decomposing the FRB dispersion measure, the contribution of the IGM gas to $\mathrm{DM}_{\IGM}$ is usually modelled as a quantity linearly tracing the distribution of the matter density field (Equation~\ref{eq:flimflam_assumption}).
In other words, we show in Section~\ref{sec:f_gas} that the assumption $\rho_{\mathrm{gas, IGM}} = (1+\delta_m) \bar{\rho}_{\mathrm{gas, IGM}}$ (required by Equation~\ref{eq:flimflam_assumption}) may not hold when jet feedback is present, such as in the \ff{} and \nj{} simulations. 
This long-range collimated jet feedback is likely to raise $\fgas$ in low-density regions relative to high-density ones. 
Overall, this could lead to a bimodal variation in the $\mathrm{DM}_{\IGM}$ estimation, introducing systematics to the FRB analysis. 

To account for the influence of jet feedback on $\mathrm{DM}_{\IGM}$, we suggest modelling a local gas fraction $\fgas$ dependent on overdensity, rather than a global value of $f_{\IGM}$. 
Figure~\ref{fig:f_gas} shows the $\fgas$ distribution by overdensity; this correction may serve as a bias term when translating $(1+\delta_m)$ into the density of electron numbers. The DM test in Figure~\ref{fig:DM_test} further warns about the risk of underestimating $\mathrm{DM}_\IGM$ following the global $\figm$ formulation detailed in Equation~\ref{eq:flimflam_assumption}. 

At the current spatial resolution of the density reconstructions in the FLIMFLAM survey (around 2 cMpc/h), the changes to $\fgas$ seem to minimally affect the DM budget across the cosmic web. 
\cite{Borrow:2020} reports that 90\% of baryons in \simba simulations with jets are spread within 3 cMpc/h around their initial Lagrangian regions (the comoving volume in the initial condition that corresponds to the halo volume at lower redshifts), extensive yet smaller than the structures defined here. 
Meanwhile, the typical spread of the no-jet runs is merely 1 cMpc/h, comparable to the largest haloes in the simulation. 
The findings affirm the use of full-density-based reconstruction treatment in FLIMFLAM, with the separation of the halo CGM and large-scale IGM as a probe of galaxy feedback. 
Given the reconstructed cell size of approximately $2$ cMpc, modelling large-scale baryonic effects as secondary is justified.
Nonetheless, the deviation of the gas fraction from unity becomes more significant when analysing structures below the megaparsec scale. 
In the upcoming FLIMFLAM data release, we will evaluate an improved density reconstruction method (Dong et al. in prep), potentially highlighting the impact of non-unity $\fgas$ for diffused IGM gas.

\section{Conclusion} \label{sec:conclusion}
In this study, we use the cosmic web classification based on eigenvalues (T-web) to examine the distribution of IGM gas in \simba with special consideration of the influence of feedback mechanisms. 
The key findings are as follows.
\begin{itemize}[leftmargin=*]
    \item On scales of the reconstructed cosmic web (typically $\gtrsim$ a few Mpc), the classification of large-scale environments via the T-web method remains largely stable against variations in feedback models. 
    Changes in these feedback models induce only minor variations in classification outcomes (see Figure~\ref{fig:class_fields}). 
    Furthermore, the associated mass and volume fractions indicate that the classification is relatively insensitive to feedback physics (see the upper panels of Figure~\ref{fig:four_in_one}).
    
    \item Analysis of the gas (lower left panel of Figure~\ref{fig:four_in_one}) and electron distribution (lower right panel of Figure~\ref{fig:four_in_one}) indicates that jet feedback is crucial in modifying their distribution, although in a relatively subtle way. 
    More baryons remain in sheets, and fewer baryons enter knots when jet feedback is active. 
    Jet feedback also effectively removes baryons from haloes, as confirmed by \cite{Khrykin:2024} and \cite{Sorini2022}. 
    \item Jet feedback minimally affects ($\lesssim$ 1\%) the partition of the IGM gas across the cosmic web structures (Figure~\ref{fig:fraction_normalize_by_IGM}).
    This indicates a limited advantage in incorporating differential baryon fractions across cosmic web structures for the FRB foreground mapping analysis (in Equation~\ref{eq:new_decomp}).
    \item By extending the definition of gas fraction $\fgas$ from individual haloes in \cite{Khrykin:2024} to voxels with $0.1$ cMpc/h size, 
    We note that variations in the $\fgas$ values are spatially correlated with physical large-scale structures, especially when jet feedback is involved (Figure~\ref{fig:f_gas_vis}).
    Analysis of the dependence of $\fgas$ on the overdensity field suggests that jet feedback causes an excess of baryon ratio in the diffuse IGM for $0.1<1+\delta_m<10$ (Figure~\ref{fig:f_gas}), mainly from the volume of filaments or knots. 
    A simple test on the dispersion measure (Figure~\ref{fig:DM_test}) suggests that the DM can be underestimated when feedback is not taken into account in the modelling, which introduces systematics in the inference of the $\figm$ parameter. 
    The underestimation occurs primarily when the FRB line of sight does not intersect any massive structures and becomes more pronounced in the runs with powerful jet feedback.
\end{itemize} 

Our results highlight the role of AGN jet feedback in shaping the baryon distribution within haloes and the IGM, as revealed by the \simba simulation and analysed through the cosmic web.

We tested the extension of the FLIMFLAM baryonic model with cosmic web classification and reported the dependence of $\fgas$ on the local IGM overdensity, which is a notable systematic to be addressed with future FRB foreground mapping projects.
Alternatively, modification of $\fgas$ might be achieved through a baryon painting neural network (e.g., \citealt{Troster:2019, Horowitz:2022, Chadayammuri:2023, Williams:2023}). 
The relationship could be represented using emulators trained with cosmological simulation datasets for machine learning applications, such as CAMELS \citep{Paco:2021CAMELS,Guo:2025}, and incorporated into the FRB foreground mapping data model (see \citealt{Paco:2021, Mohammad:2022} for associated machine learning applications).
While our analysis is limited to redshift $z=0$, it is straightforward to extrapolate to $z<0.5$ given the mild evolution of feedback in this epoch \citep{Sorini2022}. 
We leave the analysis on higher redshift to future works. 

Although our primary motivation stems from FRB foreground mapping, the broader implications extend to general FRB-based constraints on the cosmic baryon budget. 
Looking ahead, forthcoming radio surveys such as CHIME, DSA-2000, and CRAFT, combined with deep optical follow-ups from Subaru/PFS, DESI, Roman, and LSST, will enable a powerful synergy between radio transients and traditional galaxy surveys. 
This multi-wavelength approach promises to significantly advance our understanding of the baryon distribution in the low-redshift universe.

\section*{Acknowledgements}
Kavli IPMU is supported by the World Premier International Research Center Initiative (WPI), MEXT, Japan. 
This work was performed in part at the Center for Data-Driven Discovery, Kavli IPMU (WPI).
This study was supported by the Forefront Physics and Mathematics Program to Drive Transformation (FoPM), a World-leading Innovative Graduate Study (WINGS) Program, the University of Tokyo.
CD thanks Sunil Simha and Ilya Khrykin for useful discussions. 
FD would like to thank Kavli IPMU for having allowed this collaboration, and especially the \texttt{idark} team, which has been providing a quality service that was very valuable for the smooth running of this project.
KGL acknowledges support from JSPS Kakenhi grants no. JP19K14755 and JP24H00241.

%%%%%%%%%%%%%%%%%%%%%%%%%%%%%%%%%%%%%%%%%%%%%%%%%%
\section*{Data Availability}

The \simba cosmological simulation is publicly available at \href{http://simba.roe.ac.uk/}{Simba project repository}. 
The data products in this work, including the density field and cosmic web classification, will be shared upon request to the authors.

%%%%%%%%%%%%%%%%%%%% REFERENCES %%%%%%%%%%%%%%%%%%

% The best way to enter references is to use BibTeX:

\bibliographystyle{mnras}
\bibliography{example} % if your bibtex file is called example.bib

\appendix

\section{Dispersion Measure Decomposition in FLIMFLAM}
\label{appendix:FLIMFLAM_intro}

We present the FRB DM decomposition adopted in FLIMFLAM and the refined model in this appendix. 
The readers may refer to the FLIMFLAM papers \citep{Lee_2022, Huang:2024} for more details. 

Following the decomposition in Equation \ref{eq:DM_FLIMFLAM}, the foreground galaxy observations aim to find haloes along the FRB line of sight (``intervening haloes'') and also to reconstruct the large-scale density field $\rho_m$ using galaxies as biased tracers (e.g. the ARGO code used in FLIMFLAM by \citealt{Ata:2015, Ata:2017}). 
Then $\mathrm{DM}_{\mathrm{haloes}}$ is calculated with the intervening haloes assuming a radial profile (see \citealt{Khrykin:2024} for a further discussion),
while $\mathrm{DM}_{\IGM}$ is calculated following Equation~\ref{eq:flimflam_assumption} (using Equation~\ref{eq:define_dm}):
\begin{equation} \label{eq:flimflam_assumption}
    \mathrm{DM_{IGM}} = \int n_{e, \IGM} \frac{ds}{1+z} \approx \int \bar{n}_{e, \IGM} (1+\delta_m) \frac{ds}{1+z}
\end{equation}
where $\delta_m(\mathbf{x}) = \rho_m(x) / \bar{\rho}_m - 1$ is the mass overdensity, and $\bar{n}_{e, \IGM}$ is the mean electron number density in the IGM, defined in the following Equation~\ref{eq:n_igm}.
\begin{align}\label{eq:n_igm}
    \bar{n}_{e, \IGM} 
    &= f_{\IGM} \Afrac{}
\end{align}
In this model, $f_{\IGM}$ is the fraction of baryons residing in the IGM, a free parameter to be determined in FLIMFLAM's analysis. 
$m_\mathrm{H}$ and $m_\mathrm{He}$ are the atomic masses of hydrogen and helium, respectively; $X_\mathrm{H}=0.757$ and $Y_\mathrm{He} = 0.243$ are the cosmic
mass fraction of the two elements. 
Also, in Equation~\ref{eq:flimflam_assumption}, the default assumption is that the baryon-to-total-mass fraction is the cosmic value ($\rho_{\mathrm{gas, IGM}} = (1+\delta_m) \bar{\rho}_{\mathrm{gas, IGM}}$) and the IGM gas is fully ionized ($n_{e, \mathrm{gas, IGM}} \propto \rho_{\mathrm{gas, IGM}} $). 

In our refined model incorporating cosmic web (Equation~\ref{eq:new_decomp}), we consider the decomposition of DM as 
\begin{equation}\label{eq:new_decomp_appendix}
    \begin{aligned}
    \mathrm{DM}_{\mathrm{IGM}} &= \mathrm{DM}_{\mathrm{voids}} + \mathrm{DM}_{\mathrm{sheets}} + \mathrm{DM}_{\mathrm{filaments}} + \mathrm{DM}_{\mathrm{knots}}\\
     &=\sum_{j} \int_{j} f_{\IGM, j} (1+\delta_m) A(z) \frac{ds}{1+z}\\
     &= \sum_{j} f_{\IGM, j} \int_{j}  (1+\delta_m) A(z) \frac{ds}{1+z}
    \end{aligned}
\end{equation}

Where $f_{\mathrm{IGM, j}} = \frac{M_{\mathrm{IGM}}}{M_{\mathrm{baryon}}}|_{j}$, $j \in \{\mathrm{voids, sheets, filaments, knots}\}$ is the fraction of baryons in the IGM for the different cosmic web environments, while $A(z) = \Afrac{}$. As it is already revealed by \cite{Khrykin:2024} that the different feedback models will be reflected in the global $\figm$ value, we will rewrite $f_\mathrm{IGM, j}$ as
\begin{equation}\label{eq:f_igm_breakdown}
\begin{aligned}
    f_\mathrm{IGM, j} 
    &= \frac{M_{\mathrm{all, IGM}}}{M_{\mathrm{all, baryon}}}\frac{M_{\mathrm{all, baryon}}}{M_{\mathrm{j, baryon}}}\frac{M_\mathrm{j, IGM}}{M_\mathrm{all, IGM}}\\
    &= f_{\mathrm{IGM}} \frac{F_{\mathrm{j, IGM}}}{F_{\mathrm{j, baryon}}},
\end{aligned}
\end{equation}
where $F_{j,\IGM} = \frac{M_{\mathrm{j, IGM}}}{M_{\mathrm{all, IGM}}}$ and $F_{j, \mathrm{baryon}} = \frac{M_{\mathrm{j, baryon}}}{M_{\mathrm{all, baryon}}}$ are the partitions of IGM gas and baryons in each type of structure, respectively. 
The potential improvement of the model that incorporates cosmic web structures would stem from the variation in their ratio $F_{j,\IGM} / F_{j, \mathrm{baryon}}$ among the feedback models.

\section{Data Table}
\label{appendix:data_table}

In this appendix, we show the fraction data (Table \ref{table:relative_matter_part}) used in Figures~\ref{fig:four_in_one} and \ref{fig:fraction_normalize_by_IGM}.
We listed the fractions of mass, volume, gas, and electron number in the cosmic web structures (voids, sheets, filaments, and knots) for different feedback variants (\ff{}, \nx{}, \nj{}, \nagn{}, and \nf{}) in percentage. 
In parentheses, we also included the fractions when calculated only for material outside haloes. 

\begin{table*} 
\caption{
The fraction of total mass (dark matter and baryon), volume, baryonic gas and electron number in \simba feedback variants according to cosmic web classification. 
Note values inside parentheses are statistics with haloes excluded. 
}

\label{table:relative_matter_part}      % is used to refer this table in the text
\centering                          % used for centering table
\scalebox{1.0}{
\begin{tabular}{ccccc}
Total Mass Fraction   {[}\%{]} & voids & sheets & filaments & knots \\ \hline
$\ff{}$ & 8.39(11.46) & 26.10(31.99) & 39.99(39.42) & 25.52(17.14) \\
$\nx{}$ & 8.37(11.51) & 25.90(31.91) & 40.14(39.52) & 25.60(17.06) \\
$\nj{}$ & 8.26(11.94) & 25.33(32.34) & 40.10(38.96) & 26.31(16.76) \\
$\nagn{}$ & 8.25(12.41) & 25.28(33.79) & 40.04(40.40) & 26.43(13.40) \\
$\nf{}$ & 8.23(12.52) & 25.12(33.16) & 40.13(39.36) & 26.51(14.96) \\ \hline
Volume Fraction {[}\%{]} & voids & sheets & filaments & knots \\ \hline
$\ff{}$ & 33.90(33.95) & 44.06(44.11) & 20.28(20.25) & 1.76(1.69) \\
$\nx{}$ & 33.90(33.95) & 44.02(44.07) & 20.32(20.29) & 1.76(1.69) \\
$\nj{}$ & 33.85(33.91) & 43.98(44.04) & 20.41(20.36) & 1.77(1.70) \\
$\nagn{}$ & 33.84(33.89) & 43.98(44.04) & 20.41(20.38) & 1.77(1.69) \\
$\nf{}$ & 33.80(33.86) & 43.96(44.02) & 20.47(20.43) & 1.78(1.69) \\ \hline
Gas Fraction {[}\%{]} & voids & sheets & filaments & knots \\ \hline
$\ff{}$ & 9.22(10.17) & 29.88(32.21) & 40.37(42.06) & 20.53(15.56) \\
$\nx{}$ & 9.21(10.23) & 29.43(31.76) & 40.80(42.20) & 20.56(15.82) \\
$\nj{}$ & 9.28(12.10) & 27.68(33.41) & 40.48(39.88) & 22.56(14.61) \\
$\nagn{}$ & 9.39(12.62) & 27.73(34.55) & 40.25(40.74) & 22.63(12.09) \\
$\nf{}$ & 10.42(13.54) & 28.68(34.67) & 39.14(39.23) & 21.76(12.56) \\ \hline
Electron Number Fraction {[}\%{]} & voids & sheets & filaments & knots \\ \hline
$\ff{}$ & 8.20(9.12) & 28.27(30.85) & 40.68(42.87) & 22.84(17.16) \\
$\nx{}$ & 8.09(9.04) & 27.72(30.24) & 41.27(43.20) & 22.93(17.52) \\
$\nj{}$ & 8.25(10.93) & 25.81(31.78) & 40.61(40.71) & 25.33(16.58) \\
$\nagn{}$ & 8.44(11.74) & 25.74(33.21) & 40.16(41.44) & 25.65(13.62) \\
$\nf{}$ & 9.25(12.50) & 26.20(32.85) & 39.24(39.98) & 25.32(14.67)
\end{tabular}
}
\end{table*}

\section{Convergence test} \label{appendix:convergence_test}
To test the impact of box size, we show in Table~\ref{table:conv_diff_boxsize_resolution} the mass and volume fractions of the cosmic web structures for the \simba-100 and \simba-25 runs.
Both runs are equipped with full feedback physics and shares the same mass resolution as the \ff{} run. Their box sizes are 100 cMpc/h and 25 cMpc/h, respectively. 
Comparing \simba-100 with \ff{}, we find that the difference in either mass and volume fraction, between this run and the \ff{} run, is within 2\%. 
For the \simba-25, the difference in fractions is up to 5\%. 
This result is acceptable given that the smaller box should suffer from severer cosmic variance. 
The \simba-25-hi run constitutes a high-resolution simulation, characterized by a $8\times$ mass resolution in comparison to \ff{}, within a box size of 25 cMpc/h. 
A notable distinction is observed in the statistical properties of sheets and filaments. 
Specifically, the mass fraction of sheets diminished by 3\%, while the volume fraction of sheets decreased by 5\%. 
Meanwhile, the mass and volume fractions of filaments augmented by 5\% and 3\%, respectively.

The impact of $\lth$ on the T-web classification is evaluated in Table \ref{table:diff_lambda}.
For this convergence test, we stick to the smoothing scale of $2$ cMpc/h due to the spatial resolution of FLIMFLAM density reconstruction (Section \ref{subsec:class}).
Fraction values from $\lth=0.05$ to $\lth=0.40$ indicate that the variations in mass fractions are $+14\%, +4\%, -13\%, -5\%$ for voids, sheets, filaments, and knots, respectively. 
The volume fraction demonstrates greater sensitivity to the selection of $\lth$, especially for voids: the volume fraction of voids increases by approximately $25\%$ from $\lth=0.05$ to $\lth=0.40$. 
This characteristic of the T-web method was already mentioned by \cite{Forero_Romero_2009}.

\begin{table*}
\caption{
The fraction of total mass and volume in box size and mass resolution variants. 
}

\label{table:conv_diff_boxsize_resolution}      % is used to refer this table in the text
\centering                          % used for centering table
\scalebox{1.0}{
\begin{tabular}{ccccc}
Total Mass Fraction   {[}\%{]} & voids & sheets & filaments & knots \\ \hline
$\simba$-100 & 8.12 & 28.29 & 40.45 & 23.15 \\
$\simba$-25 & 7.09 & 29.23 & 42.61 & 21.06 \\
$\simba$-25-hi & 8.72 & 23.77 & 45.20 & 22.31 \\ \hline
Volume Fraction {[}\%{]} & voids & sheets & filaments & knots \\ \hline
$\simba$-100 & 32.05 & 46.74 & 19.86 & 1.35 \\
$\simba$-25 & 29.70 & 48.03 & 20.22 & 2.04 \\
$\simba$-25-hi & 36.04 & 38.63 & 23.46 & 1.87
\end{tabular}
}
\end{table*}

\begin{table*}
\caption{
The fraction of total mass and volume in $\lambda_{\mathrm{th}}$ variants. The smoothing scale $R_\mathrm{sm}=2$ cMpc/h was kept constant due to the spatial resolution of FLIMFLAM density reconstruction.
}\label{table:diff_lambda}
\centering                          % used for centering table
\scalebox{0.9}{
\begin{tabular}{ccccc}
Total Mass   Fraction {[}\%{]} & voids & sheets & filaments & knots \\ \hline
$\lr=(0.05,2\,\mathrm{cMpc/h})$ & 4.27 & 23.89 & 44.59 & 27.25 \\
$\lr=(0.10,2\,\mathrm{cMpc/h})$ & 6.00 & 25.15 & 42.50 & 26.34 \\
$\lr=(0.15,2\,\mathrm{cMpc/h})$ & 8.01 & 25.93 & 40.41 & 25.65 \\
$\lr=(0.20,2\,\mathrm{cMpc/h})$ & 9.99 & 26.66 & 38.35 & 25.00 \\
$\lr=(0.25,2\,\mathrm{cMpc/h})$ & 11.93 & 27.49 & 36.29 & 24.28 \\
$\lr=(0.30,2\,\mathrm{cMpc/h})$ & 14.06 & 27.66 & 34.74 & 23.53 \\
$\lr=(0.35,2\,\mathrm{cMpc/h})$ & 15.92 & 28.05 & 33.12 & 22.90 \\
$\lr=(0.40,2\,\mathrm{cMpc/h})$ & 17.73 & 28.42 & 31.73 & 22.12 \\ \hline
Volume Fraction {[}\%{]} & voids & sheets & filaments & knots \\ \hline
$\lr=(0.05,2\,\mathrm{cMpc/h})$ & 20.94 & 50.04 & 26.81 & 2.22 \\
$\lr=(0.10,2\,\mathrm{cMpc/h})$ & 26.87 & 47.64 & 23.50 & 1.99 \\
$\lr=(0.15,2\,\mathrm{cMpc/h})$ & 32.75 & 44.68 & 20.77 & 1.80 \\
$\lr=(0.20,2\,\mathrm{cMpc/h})$ & 38.35 & 41.52 & 18.50 & 1.63 \\
$\lr=(0.25,2\,\mathrm{cMpc/h})$ & 43.47 & 38.48 & 16.57 & 1.48 \\
$\lr=(0.30,2\,\mathrm{cMpc/h})$ & 47.91 & 35.84 & 14.91 & 1.34 \\
$\lr=(0.35,2\,\mathrm{cMpc/h})$ & 51.67 & 33.63 & 13.48 & 1.22 \\
$\lr=(0.40,2\,\mathrm{cMpc/h})$ & 55.01 & 31.61 & 12.27 & 1.10
\end{tabular}
}
\end{table*}

\label{lastpage}
\end{document}